\documentclass[sn-mathphys-ay]{sn-jnl}% Math and Physical Sciences Author Year Reference Style
\usepackage{graphicx}%
\usepackage{multirow}%
\usepackage{amsmath,amssymb,amsfonts}%
\usepackage{amsthm}%
\usepackage{mathrsfs}%
\usepackage[title]{appendix}%
\usepackage{xcolor}%
\usepackage{textcomp}%
\usepackage{manyfoot}%
\usepackage{booktabs}%
\usepackage{algorithm}%
\usepackage{algorithmicx}%
\usepackage{algpseudocode}%
\usepackage{listings}%
\usepackage{bm}
\usepackage{tabularx}
\usepackage{hyperref}
\usepackage{siunitx}
\usepackage{xfrac}
\usepackage{soul}

\input{linestyles.tex}
\newcommand{\sd}{\, {\rm d}}

\renewcommand{\hl}{}

\begin{document}

\title[Free surface influence in OCF]{How far does the influence of the free surface extend in turbulent open channel flow?}

%%=============================================================%%
%% GivenName	-> \fnm{Joergen W.}
%% Particle	-> \spfx{van der} -> surname prefix
%% FamilyName	-> \sur{Ploeg}
%% Suffix	-> \sfx{IV}
%% \author*[1,2]{\fnm{Joergen W.} \spfx{van der} \sur{Ploeg} 
%%  \sfx{IV}}\email{iauthor@gmail.com}
%%=============================================================%%
\author*[1]{\fnm{Christian} \sur{Bauer}}\email{christian.bauer@dlr.de}

\author[2]{\fnm{Yoshiyuki} \sur{Sakai}}\email{yoshiyuki.sakai@tum.de}
%\equalcont{These authors contributed equally to this work.}

\author[3]{\fnm{Markus} \sur{Uhlmann}}\email{markus.uhlmann@kit.edu}
%\equalcont{These authors contributed equally to this work.}

\affil*[1]{\orgdiv{Institute of Aerodynamics and Flow Technology}, \orgname{German Aerospace Center}, \orgaddress{\street{Bunsenstr. 10}, \city{G\"ottingen}, \postcode{37073}, 
%\state{Lower Saxony}, 
\country{Germany}}}

\affil[2]{\orgdiv{TUM School of Engineering and Design}, \orgname{Technical University of Munich}, \orgaddress{\street{Arcisstr. 21}, \city{Munich}, \postcode{80333}, 
%\state{Bayern}, 
\country{Germany}}}

\affil[3]{\orgdiv{Institute for Water and Environment}, \orgname{Karlsruhe Institute of Technology}, \orgaddress{\street{Kaiserstr. 12}, \city{Karlsruhe}, \postcode{76131}, 
%\state{Baden-Wuerttemberg}, 
\country{Germany}}}
%==============================================================================
%
% ==============================================================================
%\begin{abstract}
\abstract{
  Turbulent open channel flow is known to feature a multi-layer
  structure near the free surface. 
  In the present work we employ direct numerical simulations considering
  Reynolds numbers up to $\mathrm{Re}_\tau=900$ and domain sizes large
  enough ($L_x=12 \pi h$, $L_z=4 \pi h$) to faithfully capture the
  effect of very-large-scale motions in order to
  test
  the proposed scaling laws and
  ultimately
  answer the question: How far does the
  influence of the free surface extend?  
  In the region near the free surface, where fluctuation intensities of
  velocity and vorticity become highly anisotropic, we observe the
  previously documented triple-layer structure, consisting of a
  wall-normal velocity damping layer that scales with the channel height
  $h$, and two sublayers that scale with the near-surface viscous length
  scale $\ell_\mathrm{V}=\mathrm{Re}_\mathrm{b}^{-1/2}h$ and with the Kolmogorov length
  scale $\ell_\mathrm{K}=\mathrm{Re}_\mathrm{b}^{-3/4}h$, respectively. 
  The scaling laws previously proposed by Calmet and Magnaudet
  [J. Fluid. Mech. \textbf{474}, 355--378 (2003)]
  are found
  to hold with the following exceptions.
  The thin layer, where the intensity of
  surface-parallel components of the vorticity rapidly decreases to
  zero, is here found to scale with the Kolmogorov length scale
  $\ell_\mathrm{K}$ rather than with the near-surface viscous scale $\ell_\mathrm{V}$.  
  In addition, we argue that the Kolmogorov length scale is the relevant
  scale for the mean velocity gradient near the free surface. 
  Both the mean velocity gradient and 
  the fluctuation intensity of the surface-parallel component of vorticity 
  decay to zero in the Kolmogorov sublayer $\delta_\mathrm{K} \approx 20 \ell_\mathrm{K}$.
  On the other hand, the layer, where the wall-normal turbulence
  intensity decreases linearly to zero near the free surface, scales
  with $\ell_\mathrm{V}$ rather than $\ell_\mathrm{K}$ as suggested by Calmet and
  Magnaudet.  
  The corresponding near-surface viscous sublayer measures $\delta_\mathrm{V}
  \approx \ell_\mathrm{V}$.
  Importantly, the streamwise turbulence intensity profile for
  $\mathrm{Re}_\tau \ge 400$ suggests that the influence of the
  free-slip boundary penetrates essentially all the way down to the
  solid wall through the appearance of enhanced very-large-scale motions
  ($\delta_{\mathrm{SIL}}\approx h$). 
  In contrast, the layer where the surface-normal turbulence intensity
  is damped to zero is restricted to the free surface
  ($\delta_{\mathrm{NVD}}\approx 0.3h$). 
  As a consequence, the partitioning of the surface-influenced region has to be
  expanded to a four-layer structure that spans the entire channel
  height $h$. 
}
%\end{abstract}
% ==============================================================================
%
% ==============================================================================
% Uncomment for keywords
%\vspace{2pc}
%\noindent{\it Keywords}: turbulent open channel flow, direct numerical simulation, free surface
%
% Uncomment for Submitted to journal title message
%\submitto{Flow, Turbulence and Combustion}
%
% Uncomment if a separate title page is required
%\maketitle
% 
% For two-column output uncomment the next line and choose [10pt] rather than [12pt] in the \documentclass declaration
%\ioptwocol
% ==============================================================================
%
%
%
% ==============================================================================
\maketitle

\section{Introduction}
\label{sec:intro}
% ------------------------------------------------------------------------------
Plane Poiseuille flow, also known as closed channel flow (CCF), is one of the most studied canonical flows by means of direct numerical simulations \cite[DNSs,][]{Kim1987,Moser1999,Hoyas2006,Lee2015,Oberlack2022}.
The numerical domain is defined by doubly-periodic 
%boundary condition 
boundaries 
in the stream- and spanwise directions, and impermeable no-slip walls at the bottom and the top.
Conversely, less attention has been paid to open channel flow (OCF), where one of the no-slip walls is replaced by a free-slip plane, despite its direct relevance to environmental flows.
As noted by \citet{Pinelli2022}, OCF is of particular interest when studying the mass transfer across gas-liquid interfaces in oceans or engineering applications.
One of the defining characteristics of OCF is the transfer of kinetic energy from the vertical to the surface-parallel velocity components due to the damping effect of the free surface~\citep{Swean1991,Leighton1991,Handler1993,Komori1993,Pan1995}.
While the inter-component kinetic energy transfer is dominated by the small scales in the vicinity of the free surface, \citet{Pinelli2022} showed that low-speed large- and very-large-scale motions ((V)LSMs) are much more correlated with the regions of small-scale vortices than high-speed (V)LSMs. Hence, the spatial distribution of the small scales in the vicinity of the free surface is modulated by the large scales.
% ------------------------------------------------------------------------------
Most of the early OCF DNSs were restricted to small friction Reynolds numbers $\mathrm{Re}_\tau = u_\tau h/\nu$, based on friction velocity $u_\tau$, channel height $h$ and kinematic viscosity $\nu$, and/or small computational domains.
Recently, \citet{Yao2022} compared CCF and OCF by means of DNS data in computational domains of $L_x \times L_z=8\pi h \times 4 \pi h$ for Reynolds numbers up to $\mathrm{Re}_\tau = 2000$.
In agreement with experiments~\citep{Duan2020,Duan2021} and previous DNSs~\citep{Bauer2015}, they reported that VLSMs---which are a common feature of wall-bounded turbulent flows---are more energetic and larger in OCF than in CCF.
Generally, VLSMs become more energetic with increasing Reynolds number~\citep{Bauer2015,Duan2021,Yao2022}, leading to the failure of wall scaling of the streamwise intensity in CCF~\citep{Hoyas2006}.
As reported by~\cite{Hoyas2006} for CCF, the streamwise turbulence intensity scales neither in wall nor in bulk units.
% ------------------------------
To compare wall-attached motions between OCF and CCF, \citet{Gong2023} performed DNSs in extremely large domains of length $48\pi h$ for $\mathrm{Re}_\tau = 550$ and length $24 \pi h$ for $\mathrm{Re}_\tau = 1000$.
They found that wall-attached motions contribute more to the turbulent intensity in OCF than in CCF. 
% ---------------------------
In \citet{Bauer2024b}, the current authors carried out DNSs at friction Reynolds numbers up to $\mathrm{Re}_\tau \approx 900$ in large computational domains up to $L_x\times L_z=12\pi h \times 4\pi h$ and found that unlike CCF, the streamwise turbulence intensity in OCF scales with the bulk velocity for $\mathrm{Re}_\tau \gtrsim 400$.
The additional streamwise kinetic energy in OCF was attributed to larger and more intense VLSMs compared to CCF.
% ------------------------------------------------------------------------------

Aiming at the characterization of the multiple-layer structure in the vicinity of the free surface in OCF, \citet{Calmet2003} ---hereinafter referred to as CM03--- derived scaling laws for turbulence statistics based on the rapid distortion theory, which was originally applied by \citet{Hunt1978} to the shear-free flow along a flat surface. 

CM03 first deduced an integral length scale $L_\infty$ using the assumption of high-Reynolds-number equilibrium between kinetic energy transfer and dissipation rate, $\varepsilon\sim u_{iso}^3/\ell$, (where $u_{iso}=(2k/3)^{1/2}$\hl{,
$k=\langle u_i^\prime u_i^\prime \rangle/2$
is the turbulent kinetic energy,
}
and $\ell$ is the characteristic scale of the energetic eddies) and the relation $L_\infty=\ell/2$ which \citet{Hunt1978} found to be valid for homogeneous-isotropic turbulence.
Subsequently, CM03 performed a large-eddy simulation (LES) of OCF at $\mathrm{Re}_\tau=1280$ and compared their results to these scaling relations.
Based on their data, CM03 estimated the thickness of the so-called ``surface-influenced layer'' as $\delta_{\mathrm{SIL}}\approx L_\infty\approx0.2h$, in agreement with the 
rapid distortion 
theory of \citet{Hunt1978}. %The region from the free surface down to $y_{\mathrm{SIL}}=h-L_\infty$ was termed ``surface-influenced layer'', and its corresponding thickness was estimated as $\delta_{\mathrm{SIL}}\approx L_\infty\approx0.2h$.
% ---------------------------------------------------------------------------- %
CM03 further categorized the surface-influenced layer into the following sublayers in the near-surface region: a viscous sublayer 
(of wall-normal extent $\delta_\mathrm{V}$) 
which scales with $\mathrm{Re}_\infty^{-1/2}L_\infty$, and a Kolmogorov sublayer 
(of extent $\delta _K$)
which scales with $\mathrm{Re}_\infty^{-3/4}L_\infty$. Note that $\mathrm{Re}_\infty=u_{iso}L_\infty/\nu$ is the turbulent Reynolds number evaluated at the bottom of the surface-influenced region 
% [this does not work, since coords not defined
%($y_{\mathrm{SIL}}=h-\delta_\infty$).
(at a distance of $L_\infty$ from the free surface).
The data of CM03 suggested that the wall-parallel 
%rms vorticities 
vorticity magnitude 
decays 
to zero within the viscous sublayer, whereas the wall-normal 
%rms 
velocity 
magnitude 
grows linearly within the Kolmogorov sublayer.
Note that the LES of CM03 with a Reynolds number of $\mathrm{Re}_\tau=1280$ was performed in a 
comparatively 
small computational domain ($L_x \le 2 \pi h$, $L_z \le \pi h$) that is too restrictive to accommodate VLSMs \citep{Bauer2024b}.
Moreover, uncertainties from their sub-grid scale model motivates a follow-up investigation by means of DNS.
% ---------------------------------------------------------------------------- %

In the present work we are addressing the research question "How far does the influence of the free surface extend in turbulent open channel flow?" by revisiting the scaling predictions by CM03, whilst taking into account the latest findings regarding VLSMs and the bulk velocity scaling reported by~\citet{Bauer2024b}. For this purpose we analyze the OCF DNS data published in~\citet{BauerData2023} 
%with 
in 
a Reynolds number regime where VLSMs begin to play a role (up to $\mathrm{Re}_\tau=900$) and in a computational domain large enough to capture them ($L_x=12\pi h$, $L_z=4\pi h$) with a grid refinement towards the free surface sufficient to resolve the thin near-surface layers.\par
% ---------------------------------------------------------------------------- %
%
% ==============================================================================
%
%
%
% === NUM METHODS ============================================================ %
\section{Numerical Method}
\label{sec:methodology}
\begin{figure}
\centering
\includegraphics[width=.47\linewidth]{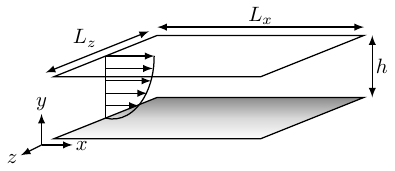}
\includegraphics[width=.47\linewidth]{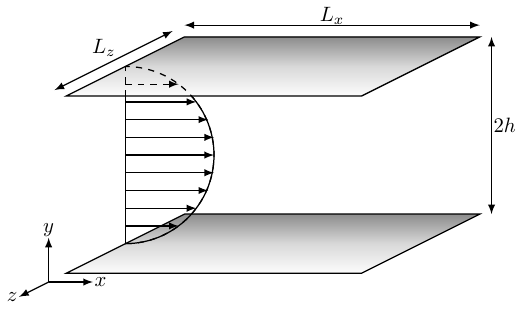}
        \caption{OCF geometry (left) and CCF geometry (right). $L_x$ and $L_z$ are the streamwise and spanwise domain length, whereas $h$ is the channel height or half-height, respectively.}
\label{fig:ocgeom}
\end{figure}

% ---------------------------------------------------------------------------- %
The numerical method is identical to the one employed by \citet{Kim1987}, where the incompressible Navier-Stokes equations are solved in their wall normal velocity-vorticity formulation. The flow variables are expanded in terms of Fourier series in the homogeneous stream- and spanwise directions, and Chebyshev polynomials in the wall normal direction, respectively. While this implies periodic boundary conditions in the stream- and the spanwise directions, a smooth no-slip wall at the bottom 
($\mathbf{u}(x,y{=}0,z){=}0)$
and a free-slip boundary condition at the top of the channel 
($\partial u(x,y{=}0,z)/\partial y{=}v(x,y{=}0,z)){=}\partial w(x,y{=}0,z)/\partial y{=}0$) 
are explicitly imposed 
in OCF 
as 
%shown
sketched 
n figure~\ref{fig:ocgeom}.
% ---------------------------------------------------------------------------- %
% ---------------------------------------------------------------------------- %
The numerical method has shown its validity in numerous CCF simulations~\citep{Kim1987,Moser1999,Jimenez1999,DelAlamo2003,DelAlamo2004}. Nevertheless, the introduction of a free-slip surface has to be regarded separately. A Chebyshev-Gauss-Lobatto grid, which is adopted in the surface-normal direction, provides a grid refinement both towards the wall and the free surface.
The necessity of the grid refinement also in the vicinity of the free-slip boundary will become obvious in the following discussion section. 
 Additional information regarding the grid refinement can be found in appendix~\ref{app:gridrefinement}.
% Figure~\ref{fig:feta} shows that the vertical spacing $\Delta y$ is kept below the value of the Kolmogorov length scale $\eta$ throughout most of the flow domain.\par
% % ---------------------------------------------------------------------------- %
% \begin{figure}% DELTAY/ETA
% \centering
% \includegraphics[]{./img/gridspacing}
% \caption{Ratio of the vertical spacing $\Delta y$ and the local Kolmogorov scale $\eta(y) = (\nu^3/\varepsilon(y))^{1/4}$ . \symba, O200; \symbb, O400; \symbc, O600; \symbd, O900. The gray line shows closed channel reference data  \cite{DelAlamo2003} at  $Re_{\tau}=550$. The dashed line indicates a reference resolution of $\pi/2$, commonly employed in DNS of homogeneous isotropic turbulence and initially stated by \citet{Jimenez1993}.}
% \label{fig:feta}
% \end{figure}
% % ---------------------------------------------------------------------------- %
\par
Our database features four different OCF and four different CCF simulation cases, summarized in table \ref{tab:cases}, where the Reynolds number and the computational domain length are varied. Statistics are either computed during runtime (turbulence intensities and spectral statistics) or from a series of instantaneous flow realizations (three-dimensional velocity correlations).
% ---------------------------------------------------------------------------- %
\begin{table} % CASES
\centering
\caption{Turbulent channel flow simulation cases as published by \cite{BauerData2023,Bauer2024b}. 
$L_x$ and $L_z$ are the stream- and spanwise domain length.
$N_x$, $N_y$ and $N_z$ are the number of grid points with respect to the streamwise, wall-normal and radial direction; $\Delta x^+$, the streamwise grid spacing; $\Delta z^+$, the spanwise grid spacing; $\Delta y^+_{\mathrm{max}}$, the maximum grid spacing in the wall-normal direction. Leading ``O'' indicates OCF cases, ``C'' CCF cases.
$\Delta T$ is the temporal averaging interval for turbulent statistics.
%$L_x/h=12\pi$ and $L_z/h=4\pi$ are the streamwise and spanwise domain lengths for all cases.
%For OCF cases: $\ell_\mathrm{V}$ and $\ell_\mathrm{K}$ are the near-surface viscous and Kolmogorov length scale, respectively.
}
%% ----------------------------------------------------------
%\begin{tabularx}{1.00\linewidth}{Xccccccccccc}
%\hline\hline
%	case & $\mathrm{Re}_{\tau}$ & $\mathrm{Re}_\mathrm{b}$  & $N_x$ &  $N_y$  & $N_z$  & $\Delta x^+$ &  $\Delta z^+$ & $\Delta y^+_{max}$ & $\Delta T u_\mathrm{b}/h$ & $\sfrac{\ell_\mathrm{V}}{h}$ & $\sfrac{\ell_\mathrm{K}}{h}$\\
%\hline
%O200 & 200 &  3170 &  768 & 129 &  512 &  9.8 & 4.9 & 2.5 & 8660 & 0.0178 & 0.0024\\
%O400 & 399 &  6969 & 1536 & 193 & 1024 &  9.8 & 4.9 & 3.3 & 1925 & 0.0120 & 0.0013\\
%O600 & 596 & 11047 & 1536 & 257 & 1536 & 14.6 & 4.8 & 3.7 & 1460 & 0.0095 & 0.0009\\
%O900 & 895 & 17512 & 3072 & 385 & 2048 & 11.0 & 5.5 & 3.7 & 1054 & 0.0076 & 0.0007\\
%C200 & 200 &  3170 &  768 & 129 &  512 &  9.8 & 4.9 & 4.9 & 8600 & & \\
%C400 & 397 &  6969 & 1536 & 193 & 1024 &  9.8 & 4.9 & 6.5 & 3260 & & \\
%C600 & 593 & 11047 & 1536 & 257 & 1536 & 14.6 & 4.8 & 7.3 & 1757 & & \\
%C900 & 889 & 17512 & 3072 & 385 & 2048 & 11.0 & 5.5 & 7.3 & 1013 & & \\
%\hline\hline
%\end{tabularx}
% ----------------------------------------------------------
 \begin{tabularx}{1.00\linewidth}{Xccccccccccc}
 \hline\hline
 case & $\mathrm{Re}_{\tau}$ & $\mathrm{Re}_\mathrm{b}$  &  $\sfrac{L_x}{h}$   & $\sfrac{L_z}{h}$   & $N_x$ &  $N_y$  & $N_z$  & $\Delta x^+$ &  $\Delta z^+$ & $\Delta y^+_{\mathrm{max}}$ & $\Delta T u_\mathrm{b}/h$ \\
 \hline
 	O200  &  200  &  3170  &  $12\pi$  &  $4\pi$  &  768  &  129  &  512  &  9.8  &  4.9  &  2.5  & 8660\\
 	O400  &  399  &  6969  &  $12\pi$  &  $4\pi$  & 1536  &  193  & 1024  &  9.8  &  4.9  &  3.3  & 1925\\
        O600  &  596  & 11047  &  $12\pi$  &  $4\pi$  & 1536  &  257  & 1536  & 14.6  &  4.8  &  3.7  & 1460\\
        O900  &  895  & 17512  &  $12\pi$  &  $4\pi$  & 3072  &  385  & 2048  & 11.0  &  5.5  &  3.7  & 1054\\
        C200  &  200  &  3170  &  $12\pi$  &  $4\pi$  &  768  &  129  &  512  &  9.8  &  4.9  &  4.9  & 8600\\
        C400  &  397  &  6969  &  $12\pi$  &  $4\pi$  & 1536  &  193  & 1024  &  9.8  &  4.9  &  6.5  & 3260\\
        C600  &  593  & 11047  &  $12\pi$  &  $4\pi$  & 1536  &  257  & 1536  & 14.6  &  4.8  &  7.3  & 1757\\
        C900  &  889  & 17512  &  $12\pi$  &  $4\pi$  & 3072  &  385  & 2048  & 11.0  &  5.5  &  7.3  & 1013\\
 \hline\hline
 \end{tabularx}
\label{tab:cases}
\end{table}
% ---------------------------------------------------------------------------- %
Hereinafter, angular brackets denote averaging over both homogeneous directions and time. Normalization in viscous or ``wall units'' is indicated by a $+$ superscript. Corresponding scales are the friction velocity $u_\tau$ and the viscous length scale $\delta_\nu=\nu/u_\tau$. Characteristic scales for normalization in ``bulk units'' on the other hand are the bulk velocity $u_\mathrm{b}$ and the channel height $h$. Hence, the bulk Reynolds number is defined as $\mathrm{Re}_\mathrm{b} = u_\mathrm{b} h / \nu$.
% ============================================================================ %
%
%
%
% ============================================================================ %
\section{Results}
\label{sec:results}
% ---------------------------------------------------------------------------- %

% ---------------------------------------------------------------------------- %
%However, CM03 also argued that $Re_\infty \sim Re_\tau$, since the integral length scale $L_\infty$ is proportional to the channel height $h$ and $u_{iso}=\mathcal{O}(u_\tau)$ at the bottom of the surface-influenced layer. Consequently, we 
%
%\marginpar{\tiny\MU{"we" or "they"? clarify}}
%
%define the near-surface length scales based on the friction Reynolds number $Re_\tau$ and the channel height $h$, which are more accessible parameters in wall bounded turbulence than $Re_\infty$ and $L_\infty$. Hence we use
As stated in section~\ref{sec:intro}, CM03 defined the near-surface length scales based on 
a turbulent Reynolds number 
$\mathrm{Re}_\infty$ 
which characterizes the ``turbulence below the surface-influenced region'' (CM03). 
However, since 
in channel flow 
$\mathrm{Re}_\infty$ is closely linked to the bulk Reynolds number $\mathrm{Re}_\mathrm{b}$, herein we define the near-surface length scales based on $\mathrm{Re}_\mathrm{b}$ and on the channel height $h$, which are 
%more 
readily 
accessible parameters in wall-bounded turbulence. 
\hl{
$\mathrm{Re}_\infty$, on the other hand, is computed from $u_{iso}$ at the bottom of the surface-influenced layer, which supposedly corresponds to the most isotropic region of the flow.
However, our OCF data show that the flow becomes less isotropic at higher Reynolds numbers due to VLSMs, which weakens the isotropy assumption. 
Moreover, the choice of the wall-normal location where $u_{iso}$ is computed is not trivial, as for instance the wall-normal coordinate of the most isotropic region might vary with the 
Reynolds number.
Consequently, in OCF $\mathrm{Re}_\mathrm{b}$ is a more robust measure than $\mathrm{Re}_\infty$, since it is an integral quantity.
 }
%than $\mathrm{Re}_\infty$ and $L_\infty$. 
%Hence we use henceforth
Henceforth we use the following definitions: 
% ---------------------------------------------------------------------------- %
\begin{eqnarray}
\ell_\mathrm{V}=\mathrm{Re}_\mathrm{b}^{-1/2}h, \label{eq:vislayer}\\
\ell_\mathrm{K}=\mathrm{Re}_\mathrm{b}^{-3/4}h, \label{eq:kollayer}
\end{eqnarray}
% ---------------------------------------------------------------------------- %
as characteristic length scales 
%to determine 
for 
the wall-normal extent of the near-surface viscous and Kolmogorov sublayers, respectively.
\hl{
Note that what is called the viscous sublayer near the free surface differs from Prandtl's viscous sublayer near the solid wall. Although viscous effects are much weaker near the free surface than near the solid wall, they also play a significant role in the vicinity of the free surface.
}
\par
% ---------------------------------------------------------------------------- %
%
% --- mean velocity profile ------------------------------------------------- %
\subsection{Mean velocity scaling}
\label{ss:umean}
\begin{figure} % d<u>/dy
\centering
\includegraphics[]{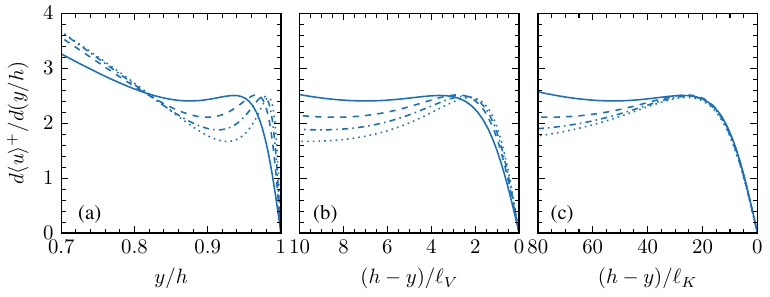}
\caption{Surface-normal derivative of the mean streamwise velocity profile $d\langle u \rangle^+ /d(y/h)$ in the vicinity of the free surface. 
Solid lines, O200; dashed lines, O400; dashed-dotted lines, O600; dotted lines, O900. 
The distance from the free surface is normalized with (a) the channel height $h$; (b) the near-surface viscous length scale $\ell_\mathrm{V}$; (c) the near-surface Kolmogorov length scale $\ell_\mathrm{K}$. }
\label{fig:dudy}
\end{figure}

% ------------------------------------------------------------------------------
In order to adapt to the free-slip boundary condition 
%($\partial u/\partial y = \partial w/\partial y = v = 0$), 
the surface-normal derivatives of the surface-parallel velocities decay to zero when approaching the free surface.
%Moreover, 
\cite{Hunt1984} {suggested that the mean velocity gradient $\sd \langle u \rangle / \sd y$ 
%stays 
remains 
approximately zero within a thin viscous sublayer near the free surface.
Nevertheless, to the knowledge of the 
present 
authors, the scaling of the mean velocity gradient near the free surface has not been investigated to date.}
Therefore, we present the profile of the surface-normal derivative of the mean streamwise velocity 
in OCF 
in figure~\ref{fig:dudy}.
%for the different Reynolds number OCF cases.
% ---------------------------------------------------------------------------- %
% ---------------------------------------------------------------------------- %
Here, the distance from the free surface is normalized with different length scales, i.e. the channel height $h$ (figure~\ref{fig:dudy}a), the near-surface viscous length scale $\ell_\mathrm{V}$ (figure~\ref{fig:dudy}b), and the near-surface Kolmogorov length scale $\ell_\mathrm{K}$ (figure~\ref{fig:dudy}c).
It can be seen that the streamwise velocity derivative profiles in the immediate vicinity of the free surface collapse well when the surface-normal coordinate is normalized with the near-surface Kolmogorov length scale $\ell_\mathrm{K}$ (figure~\ref{fig:dudy}c).
Moreover, the layer thickness where the mean velocity derivative rapidly decays to zero measures approximately $\delta_\mathrm{K} \approx20\ell_\mathrm{K}$. 
Note that unlike anticipated by \cite{Hunt1984}, a region where $\sd \langle u \rangle / \sd y$ stays approximately zero when moving away from free surface does not exist.
Finally, the observed scaling clearly shows a necessity for fine  grid resolutions near the free surface when performing OCF DNSs. We will return to this topic in section~\ref{ss:layer}
\par

% ---------------------------------------------------------------------------- %
%
% --- turbulence intensities ------------------------------------------------- %
\subsection{Turbulence intensity scaling}
\label{ss:urms}
\begin{figure} % URMS COMPARE
\centering
\includegraphics[width=\linewidth]{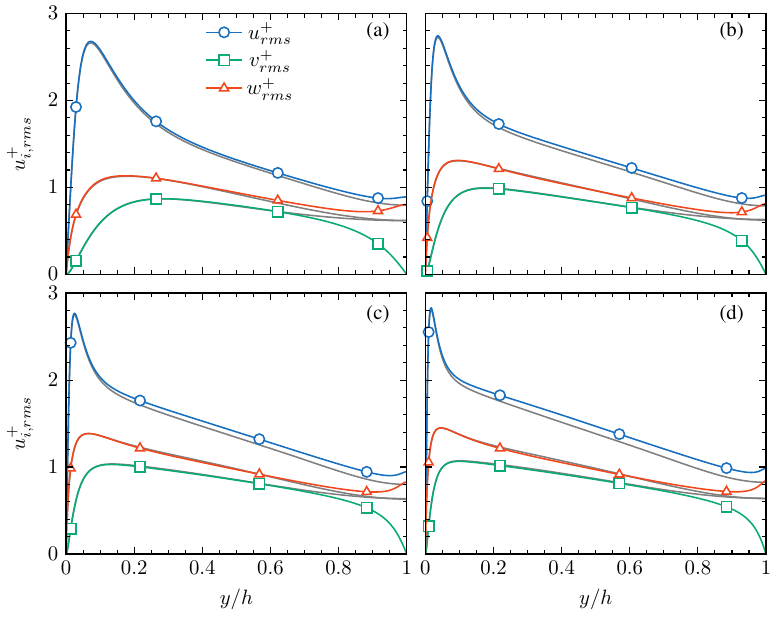}
\caption{Turbulence intensities: \symbx, $u_{rms}$; \symby, $v_{rms}$; \symbz, $w_{rms}$. Gray lines indicate CCF data, normalized by $u_{\tau}$. (a) O200, C200; (b) O400, C400; (c) O600, C600; (d) O900, C900.}
\label{fig:urmscomp}
\end{figure}

% ------------------------------------------------------------------------------
The extent of the surface-influenced layer can be estimated by comparing turbulence intensities from OCF with their closed channel counterpart, as presented in figure~\ref{fig:urmscomp}.
% ---------------------------------------------------------------------------- %
% ---------------------------------------------------------------------------- %
Note that the OCF domain upper limit at $y/h$=1 corresponds to the channel centerline of the closed channel configuration.
The spanwise and wall-normal profiles collapse fairly well with CCF results in the vicinity of the (no-slip) wall and throughout most of the flow domain.
There is, however, a noticeable discrepancy between the profiles for $y/h \rightarrow 1$ (corresponding to the centerline in CCF and to the free-surface in OCF), which is caused by the impermeable boundary
in the latter case.
The impermeability condition in OCF enforces the vertical component of the fluctuating velocity to drop to zero for $y/h \rightarrow 1$, while its energy is transferred to the surface-parallel components.
It can be seen from figure~\ref{fig:urmscomp} that the majority of the turbulence intensity $v_{rms}$ is redistributed to the spanwise direction, whereas the streamwise component increases less significantly.
This is in agreement with results obtained from numerical studies of low Reynolds number OCF~\citep{Swean1991, Komori1993, Borue1995, Nagaosa1999,Nagaosa2003,Nagaosa2012}.
In \citet{Handler1993} it was found that the pressure-strain correlation term is the key contributor to the inter-component energy transfer below the free surface.
Correspondingly, we have confirmed the role of the pressure-strain correlation for the present data (figure omitted).
%We will return to this point in section \ref{sec:surface}.

% ---------------------------------------------------------------------------- %

As already discussed above, previous authors have defined a ``surface-influenced layer'' as the region where the mean-velocity profiles in OCF deviate from their CCF counterparts (CM03). Figure~\ref{fig:urmscomp} confirms that the thickness of this layer scales with the channel height $h$, and that its extent can be quantified as $\delta_{\mathrm{NVD}}\approx0.3h$. Note that we deliberately use the new symbol $\delta_{\mathrm{NVD}}$ (for ``normal velocity damping layer'') instead of the denomination ``surface-influenced layer'' used by others, as the latter region will be shown to extend much further than previously assumed. As a consequence the commonly invoked three-layer structure of OCF will be extended to a four-layer structure.

\begin{figure} % U'3D
\centering
\includegraphics[width=\linewidth]{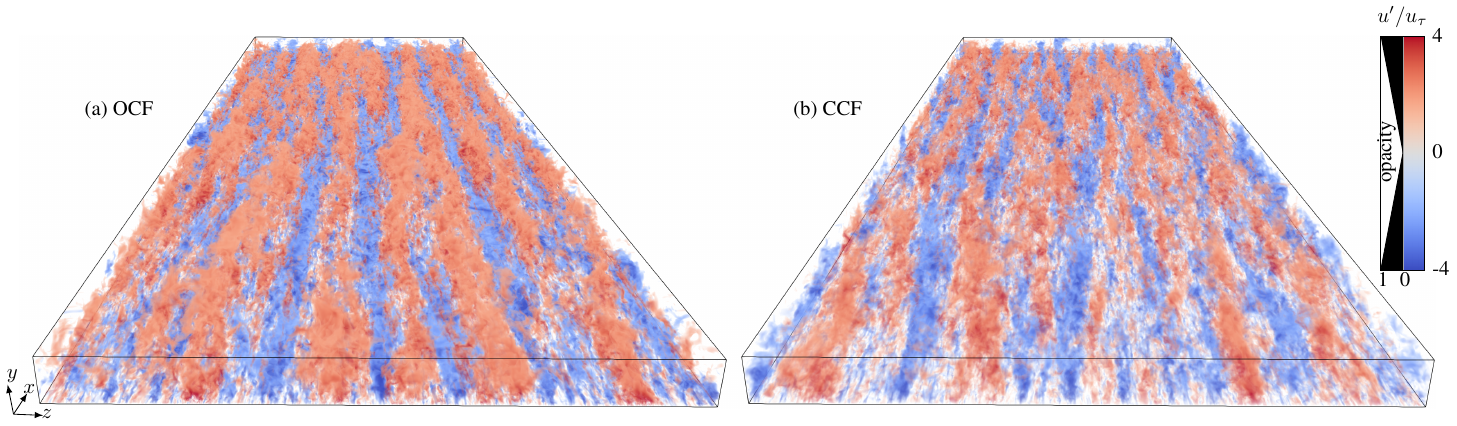}
\caption{%
%Iso-volumes of streamwise fluctuating velocity 
Volume rendering of the fluctuation of the streamwise velocity component (with respect to its streamwise and spanwise average), 
normalized in wall units $u^\prime/u_\tau$ in an instantaneous flow field for O900 (a) and C900 (b).}
\label{fig:up3d}
\end{figure}

% ---------------------------------------------------------------------------- %
We further observe in figure~\ref{fig:urmscomp} that the streamwise turbulence intensity in OCF shifts towards higher values for $y/h\gtrsim 0.1$ and $\mathrm{Re}_\tau \ge 400$ when compared with CCF values.
The increase of streamwise turbulence intensity throughout most of the flow domain is presumably caused by VLSMs appearing at Reynolds number as low as 400 in OCF.
%(see also appendix~\ref{app:vlsm}).
% ---------------------------------------------------------------------------- %
VLSMs are visible in the instantaneous streamwise 
%fluctuating 
velocity 
component 
for OCF and CCF, as depicted in figure~\ref{fig:up3d} for $\mathrm{Re}_\tau=900$.
% ---------------------------------------------------------------------------- %
% ---------------------------------------------------------------------------- %
Figure~\ref{fig:up3d} shows that VLSMs in OCF are more intense and longer in 
the 
$x$-direction than corresponding CCF structures. 
This instantaneous observation is supported by the comparison of the statistical footprint of VLSMs between OCF and CCF in appendix~\ref{app:vlsm}.\par

\begin{figure} % URMS AREA
\centering
\includegraphics[]{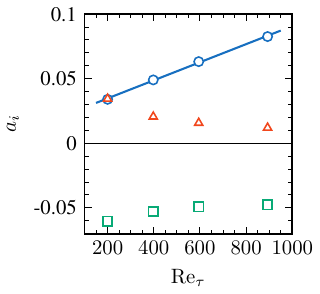}
        \caption{Integrated difference between OCF and CCF turbulence intensities: \symbxx, $a_{x}$; \symbyy, $a_{y}$; \symbzz, $a_{z}$. The blue line indicates the linear scaling law $a_x=0.0208+7\cdot10^{-5}\mathrm{Re}_\tau$.}
\label{fig:urmsarea}
\end{figure}

% ---------------------------------------------------------------------------- %
% Moreover, as we reported in \cite{Bauer2024b}, the streamwise turbulence intensity in OCF appears to scale with the bulk velocity $u_\mathrm{b}$ for $\mathrm{Re}_\tau \gtrsim 400$.
% ---------------------------------------------------------------------------- %
In order to quantify the increase in the OCF streamwise turbulence intensity 
% profile
observed in figure~\ref{fig:urmscomp},  
the integrated difference between the OCF and CCF turbulence intensity profiles is computed as follows,
% ---------------------------------------------------------------------------- %
\begin{equation}
        a_i = \int_{0}^h u^O_{i,rms} \sd y/(u^O_\tau h) - \int_{0}^h u^C_{i,rms} \sd y/(u^C_\tau h), \label{eq:urmsarea}
\end{equation}
% ---------------------------------------------------------------------------- %
where $u^O_{i,rms}$ and $u^C_{i,rms}$ denote the $i$-component of the OCF and CCF turbulence intensity, respectively, $u^O_\tau$ and $u^C_\tau$ are the OCF and CCF friction velocities (see figure~\ref{fig:ocgeom} for the definitions of $h$).
Figure~\ref{fig:urmsarea} displays the evolution of the integral quantity $a_i$ as a function of the friction Reynolds number.
% ---------------------------------------------------------------------------- %
% ---------------------------------------------------------------------------- %
While the streamwise component $a_x$ increases monotonically for the range of Reynolds numbers considered, the 
%spanwise and wall-normal 
wall-normal and spanwise
components ($a_y$, $a_z$) appear to settle at constant values for higher Reynolds numbers.
%In particular, the data presented here suggests that $a_y$ and $a_z$ settle at a constant value for higher Reynolds numbers, while $a_x$ increases linearly.
As shown in figure~\ref{fig:urmsarea}, the following linear fit,
% ---------------------------------------------------------------------------- %
\begin{equation}
        a_x = 0.0208 + 7\cdot10^{-5}\mathrm{Re}_\tau, \label{eq:urmsareafit}
\end{equation}
% ---------------------------------------------------------------------------- %
describes the data corresponding to the streamwise component reasonably well.

\begin{figure} % URMS BULK SCALING
\centering
\includegraphics[width=\linewidth]{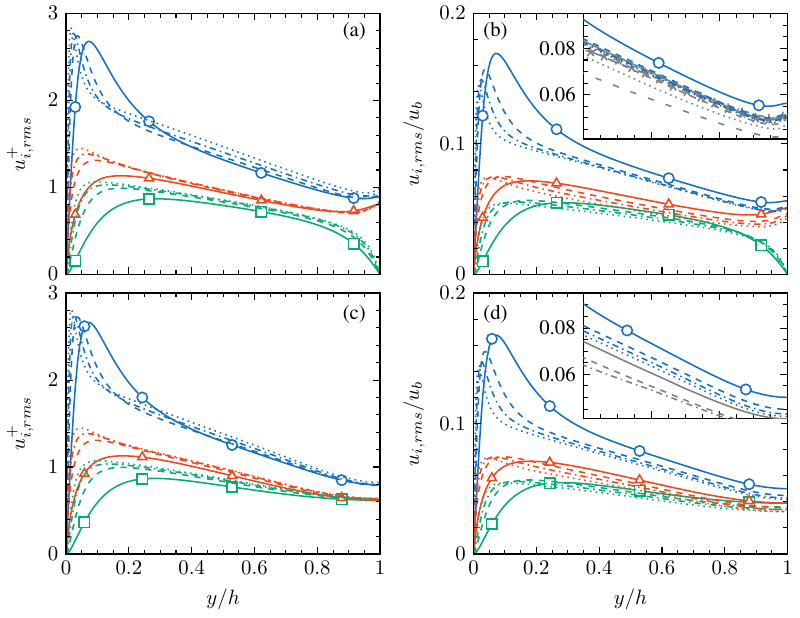}
 \caption{Turbulence intensities normalized by $u_\tau$ (a,c) and $u_\mathrm{b}$ (b,d) as function of the distance from the wall $y/h$ for open channel (a,b) and closed channel configuration (c,d): \symbx, $u_{rms}$; \symby, $v_{rms}$; \symbz, $w_{rms}$. (a,b) solid lines, O200; dashed lines, O400; dashed-dotted lines, O600; dotted lines, O900 \citep{Bauer2024b}. (c,d) solid lines, C200; dashed lines, C400; dashed-dotted lines, C600; dotted lines, C900 \citep{Bauer2024b}. The insets in (b,d) show a zoom for the streamwise turbulence intensity component.
The gray lines in (b) indicate OCF DNS data from \cite{Gong2023} at $\mathrm{Re}_\tau=550$ (\grayline) and from \citet{Pirozzoli2023} at $\mathrm{Re}_\tau=1000$ (\graydashed), 2000 (\graydashdotted), 3000 (\graydotted), 6000 (\grayloosedash).
%The gray symbols in (b) indicate profiles from OCF measurements by \cite{Duan2020,Duan2021} at $\mathrm{Re}_\tau=614$ (\graycircle), $\mathrm{Re}_\tau=1030$ (\graysquare), $\mathrm{Re}_\tau=1508$ (\graytriangle), $\mathrm{Re}_\tau=1903$ (\graydiamond), $\mathrm{Re}_\tau=2407$ (\graystar).
The symbols (\graystar) in (b) indicate a profile from OCF measurements by \citet{Duan2020,Duan2021} at $\mathrm{Re}_\tau=2407$.
The gray lines in (d) indicate CCF DNS data at $\mathrm{Re}_\tau=2003$ \citep[\grayline,][]{Hoyas2006}, $\mathrm{Re}_\tau=5186$ \citep[\graydashed,][]{Lee2015}, $\mathrm{Re}_\tau=10049$ \citep[\graydashdotted,][]{Oberlack2022}.
}
\label{fig:urms}
\end{figure}

%In 
From 
figure~\ref{fig:urmscomp} 
%it can be seen 
it was obvious 
that differences between OCF and CCF regarding the cross-stream velocity components 
are 
restricted to a layer near $y/h=1$, 
whereas the streamwise turbulence intensity differs throughout most of the channel height (for $y/h\gtrsim 0.1$).
% ---------------------------------------------------------------------------- %
The latter observation leads to a different scaling behavior of the turbulence intensities as presented in figure~\ref{fig:urms}.
% ---------------------------------------------------------------------------- %
% ---------------------------------------------------------------------------- %
While the spanwise and wall-normal contributions to the turbulence intensity scale with the friction velocity $u_\tau$ in the bulk region of the flow for $\mathrm{Re}_\tau \ge 400$ for both flow configurations (figure \ref{fig:urms}a,c), the streamwise contribution differs.
Figure \ref{fig:urms}(b) shows that the streamwise turbulence intensity in open channel flow appears to scale with the bulk velocity $u_\mathrm{b}$ for $\mathrm{Re}_\tau \ge 400$ and $y/h \gtrsim 0.3$, whereas in the closed channel case it neither scales with $u_\tau$ nor with $u_\mathrm{b}$ (figure \ref{fig:urms}c,d), as reported in \citet{Bauer2024b}.
% ---------------------------------------------------------------------------- %
Note that the described scaling behavior is only observed for domain sizes $L_x \ge 12 \pi h$ \citep{Bauer2024b}.
%Therefore, the discrepancy between the observed scaling and the high Reynolds number $u_{rms}$ profiles by \cite{Pirozzoli2023} ($\mathrm{Re}_\tau \ge 3000$) can be attributed to their box size of $L_x = 6 \pi h$.
This explains why data at high Reynolds number computed in short computational domains \citep[e.g.][cf.\ their figure 5]{Pirozzoli2023} does not exhibit the scaling of $u_{rms}$ with $u_\mathrm{b}$ 
observed here.

\begin{figure} % UU SCALING
\centering
\includegraphics[width=\linewidth]{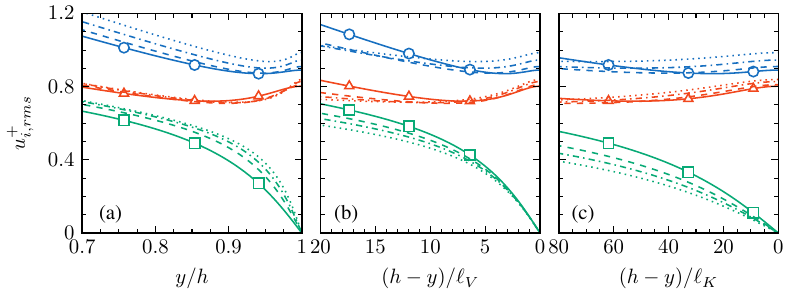}
\caption{Turbulence intensities normalized in wall units as function of distance from free surface: \symbx, $u^+_{rms}$; \symby, $v^+_{rms}$; \symbz, $w^+_{rms}$; solid lines, O200; dashed lines, O400; dashed-dotted lines, O600; dotted lines, O900. The distance from the free surface is normalized with (a) the channel height $h$; (b) the near-surface viscous length scale $\ell_\mathrm{V}$; (c) the near-surface Kolmogorov length scale $\ell_\mathrm{K}$.}
\label{fig:uuscaling}
\end{figure}

\begin{figure} % dvrms/dy 
\centering
\includegraphics[]{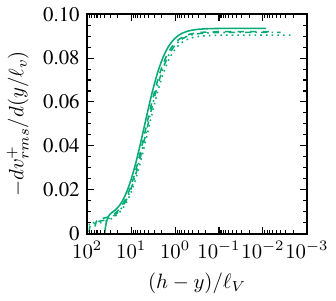}
\caption{Surface-normal derivative of the surface-normal turbulence intensity in the vicinity of the free surface $-d v^+_{rms}/d(y/\ell_\mathrm{V})$. Solid lines, O200; dashed lines, O400; dashed-dotted lines, O600; dotted lines, O900.}
\label{fig:dvrmsdy}
\end{figure}

%--------------------------------------------------------------------- %
Next we turn to the scaling of the turbulence intensities in the different surface layers in OCF.
% ---------------------------------------------------------------------------- %
Figure \ref{fig:uuscaling} shows turbulence intensity profiles normalized in wall units, plotted as a function of the distance from the free surface normalized with various characteristic length scales ($h$, $\ell_\mathrm{V}$, $\ell_\mathrm{K}$).
% ---------------------------------------------------------------------------- %
% ---------------------------------------------------------------------------- %
%Figure \ref{fig:uuscaling}(b) illustrates the surface-normal coordinate normalized by the viscous length scale $\ell_\mathrm{V}$ and it indicates that a major part ($\sim 50\%$) of the energy drop in the vertical component takes place within the thin viscous sublayer.
It can be seen that scaling $y$ with $\ell_\mathrm{V}$ provides the best collapse of the wall-normal component $v^+_{rms}$ which features a steep drop to zero within the thin viscous sublayer.
The surface-normal derivative of the surface normal turbulence intensity is presented in figure~\ref{fig:dvrmsdy}. In the immediate vicinity of the free surface ($(h-y)/\ell_\mathrm{V}\lesssim 1$) the wall-normal velocity grows linearly when moving away from the free surface, with a slope of approximately 0.09.
Therefore, the viscous sublayer is estimated as $\delta_\mathrm{V} \approx  \ell_\mathrm{V}$.
% ---------------------------------------------------------------------------- %
% ---------------------------------------------------------------------------- %
Although the region where $\partial v_{rms}/\partial y$ is constant is often referred to as the Kolmogorov sublayer~\citep[CM03]{Brumley1988}, we find that the layer scales clearly with the near-surface viscous scale $\ell_\mathrm{V}$ rather than with the Kolmogorov scale $\ell_\mathrm{K}$ (compare figure \ref{fig:uuscaling}b and c).\par

% ---------------------------------------------------------------------------- %
%
% --- rms vorticities ----------------------------------------------------------
\subsection{Vorticity scaling}
\label{ss:omega}
% ------------------------------------------------------------------------------
We observe that the intensity of vorticity fluctuations, $\langle\omega_\alpha^\prime\omega_\alpha^\prime\rangle^{1/2}$ (no summation over Greek indices), 
in OCF 
collapses well with 
corresponding 
closed channel flow profiles over most of the channel height, except for the region very close to the free surface (figure omitted).
The profile of the wall-normal vorticity component in OCF already starts to deviate from the CCF profile approximately at the bottom of the normal velocity damping layer $\delta_{\mathrm{NVD}}$.

%Then, in 
In 
the vicinity of the free surface, impermeability and free-slip boundary conditions influence all vorticity components, and for $y=h$ the vorticity vector simplifies as follows
% ---------------------------------------------------------------------------- %
\begin{equation}
\boldsymbol{\omega}(x,y=h,z)=\nabla \times \textbf{u}(x,y=h,z)=(0,\omega_{y,top},0), \label{eq:vorttop}
\end{equation}
% ---------------------------------------------------------------------------- %
while in the closed channel flow case the components show almost purely isotropic behavior at the channel center, i.e. $\omega_{x,rms}(y/h{=}1){\approx}\omega_{y,rms}(y/h{=}1){\approx}\omega_{z,rms}(y/h{=}1)$.
Note that this equality among the fluctuation intensity of the vorticity components at the channel center in CCF is progressively approached with the Reynolds number
(figure omitted). 
%\MU{Citation?}\CB{Not at hand, we see it from our data, too, maybe 
%(figure omitted)?}

\begin{figure} % OMEGA SCALING
\centering
\includegraphics[width=\linewidth]{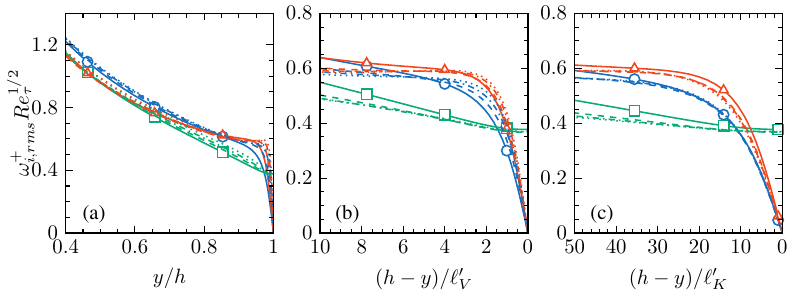}
\caption{Vorticity fluctuation intensity near the free surface, normalized by $u_\tau^2/\nu \mathrm{Re}_\tau^{-1/2}$. \symbx, $\omega_{x,rms}$; \symby, $\omega_{y,rms}$; \symbz, $\omega_{z,rms}$; solid lines, O200; dashed lines, O400, dashed dotted lines, O600; dotted lines, O900. The distance from the free surface is normalized with (a) the channel height $h$; (b) the near-surface viscous length scale $\ell_\mathrm{V}$; (c) the near-surface Kolmogorov length scale $\ell_\mathrm{K}$.}
\label{fig:omegascaling}
\end{figure}

%--------------------------------------------------------------------- %
In \citet{Antonia1996} and \citet{Panton2009} it was shown that the vorticity fluctuation intensity in the outer layer of wall-bounded turbulent flows scales with the Kolmogorov time scale, 
$\tau_\eta=(\nu/\varepsilon)^{1/2}=(\nu h/u_\tau^3)^{1/2}$ (assuming that the dissipation rate $\varepsilon$ scales with $u_\tau$ and $h$), 
leading to 
$\omega_{i,rms}{\sim}\sqrt{u^3_\tau /(\nu h) }$.
%, as dissipation is closely linked to %the vortical motions. 
%intense vortical motion. 
%
%\MU{(previous sentence not clear to me!)} 
CM03 found the same scaling to be true for the vorticity fluctuation intensity in open channel flow up to the free surface and, thus, suggested to multiply the vorticity fluctuation intensity with $\mathrm{Re}_\tau^{1/2}$ after being normalized in wall units in order to collapse data at different Reynolds numbers.
Applying the above normalization in the upper part of the channel domain, we indeed achieve a satisfactory collapse of the different Reynolds number profiles away from the wall, as shown in figure \ref{fig:omegascaling}.
It can be seen that the channel height $h$ is the characteristic length scale away both from the wall and the free surface (cf. figure \ref{fig:omegascaling}a).
% ---------------------------------------------------------------------------- %
% ---------------------------------------------------------------------------- %
Contrarily, the Kolmogorov scale $\ell_{K}$ is the relevant scale in the direct vicinity of the free surface, where the surface-parallel vorticity components decay to zero, as shown in figure \ref{fig:omegascaling}(c).
In agreement with observations from the surface-normal derivative of the mean streamwise velocity (see section~\ref{ss:umean}), the surface-parallel vorticity damping layer, where the vorticity fluctuation intensity becomes highly anisotropic as the wall-parallel vorticities rapidly decrease to zero, exhibits a thickness of approximately 20$\ell_\mathrm{K}$ (figure~\ref{fig:omegascaling}c), which is consistent with the thickness of the Kolmogorov sublayer $\delta_\mathrm{K}$.
This finding is in disagreement with CM03, who argued that this layer scales with $\ell_\mathrm{V}$ which provides an arguably less convincing collapse of our data (compare figure~\ref{fig:omegascaling}b and \ref{fig:omegascaling}c).\par
It should be recalled that CM03 did not have access to fully resolved DNS of open channel flow up to $\mathrm{Re}_\tau=900$.
Their study was limited to the comparison of LES data at $\mathrm{Re}_\tau=1280$ with only one low Reynolds number DNS at $\mathrm{Re}_\tau=134$ from \citet{Handler1993}, both of them obtained in small computational boxes and rather coarse resolutions.
A further source of uncertainty in the CM03 study is the use of a subgrid-stress model in the context of LES.\par
\begin{figure} % OMEGA SCALING
\centering
\includegraphics[width=\linewidth]{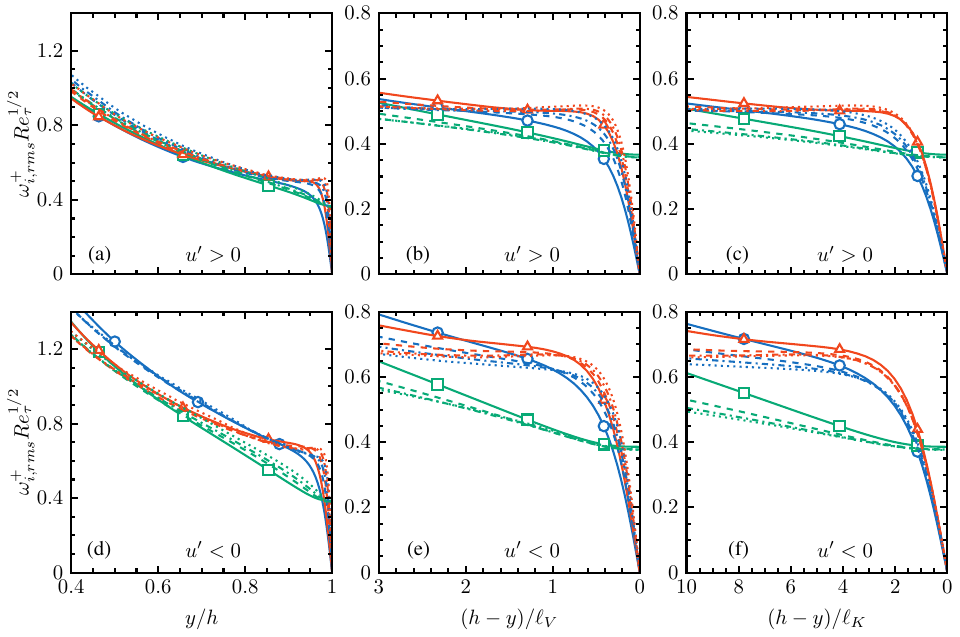}
        \caption{Conditional average of vorticity fluctuation intensity near the free surface, normalized by $u_\tau^2/\nu \mathrm{Re}_\tau^{-1/2}$. Legend see figure~\ref{fig:omegascaling}. (a,b,c) $u'>0$; (d,e,f) $u'<0$. The distance from the free surface is normalized with (a,d) the channel height $h$; (b,e) the near-surface viscous length scale $\ell_\mathrm{V}$; (c,f) the near-surface Kolmogorov length scale $\ell_\mathrm{K}$.}
\label{fig:omegacond}
\end{figure}

%{\color{red} YS: I'm wondering for the sake of CCF vs. OCF comparison, if we should mention about Hoyas \& Jimenez (PoF, 2008), in which they shows by means of the enstrophy pre-multiplied spectra (see their FIG. 5d) that "long, wide and think viscous layers have to form near the (no-slip) wall to satisfy (the no-slip boundary condition)". CB: Looking into it ...\par}
% ---------------------------------------------------------------------------- %
As pointed out by \citet{Pinelli2022}, the spatial distribution of small-scale vortices in the vicinity of the free-slip boundary is influenced by large-scale motions 
%predominantly consisting of 
in the form of 
large-scale velocity streaks.
Figure~\ref{fig:omegacond} presents the vorticity fluctuation intensities near the free-slip boundary, conditionally averaged with respect to being located within high-speed ($u'>0$) or low-speed regions ($u'<0$).
% ---------------------------------------------------------------------------- %
% ---------------------------------------------------------------------------- %
In agreement with \citet{Pinelli2022}---who found the majority of small-scale vortices in the vicinity of the free-slip boundary to be located within large-scale low-speed streaks---the vorticity fluctuation intensity is larger in low-speed regions (figure~\ref{fig:omegacond}d,e,f) than in high-speed regions (figure~\ref{fig:omegacond}a,b,c).
In the bulk flow region the 
intensity of the 
streamwise vorticity within low-speed structures appears to be slightly larger than 
that of 
the other two components (figure~\ref{fig:omegacond}d), whereas the 
intensity of vorticity fluctuations 
%vorticity components 
%
within high-speed structures 
%are 
is 
approximately isotropic (figure~\ref{fig:omegacond}a).
\subsection{Super-streamwise vortices (SSV)}
\label{ss:ssv}
% ------------------------------------------------------------------------------
%
Previous investigations have shown that large-scale motion in OCF
features elongated streamwise vortices associated to the
above-mentioned velocity streaks 
\citep{Zhong2016,Duan2021,scherer:21a}. The former structures span the entire channel
height and are sometimes termed ``super-streamwise vortices'' (SSV).
While small-scale vortices are directly detectable in both
instantaneous visualizations and statistical quantities (i.e.\ of the
vorticity field itself or through the $q$-criterion), 
super-streamwise vortices are not directly accessible, since they are
superimposed by a wide range of smaller scale turbulent motions with
higher intensity, and according to \citet{Zhong2016} statistical
evidence of their existence has been scarce.
%%
%As mentioned above, VLSMs in OCF have been associated with SSV spanning the entire channel height.
%However, these structures are not directly accessible in instantaneous flow field realizations and according to \cite{Zhong2016} statistical evidence of their existence has been insufficient.

\begin{figure}
\centering
\includegraphics[]{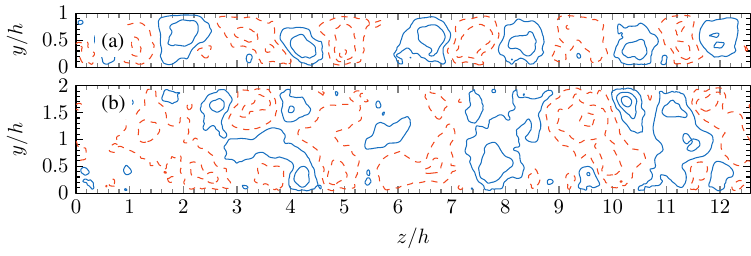}
\caption{%
Instantaneous 
iso-contours of the streamfunction of the streamwise averaged cross-sectional velocity components $\psi_{\langle v \rangle_x\langle w \rangle_x}$ in OCF (a) and CCF (b) normalized by $u_\tau h$ for $\mathrm{Re}_\tau\approx900$; Blue solid (red dashed) iso-contours indicate values of $\psi_{\langle v \rangle_x\langle w \rangle_x}=+(-)0.01$ to $+(-)0.2$ with an increment of $0.02$.}
\label{fig:psi_inst}
\end{figure}

\begin{figure}
\centering
\includegraphics[]{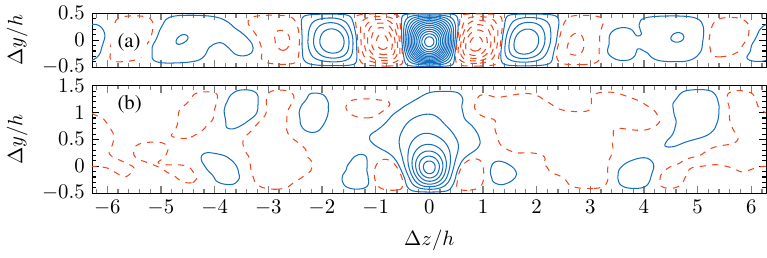}
\includegraphics[width=\linewidth]{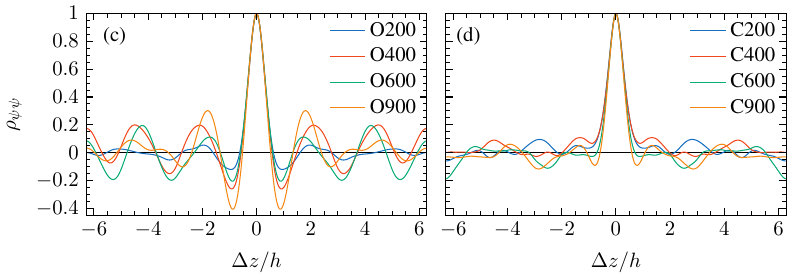}
\caption{Two-point correlation of the streamfunction of the streamwise averaged crosssectional velocity components $R_{\psi\psi}$ with $\psi=\psi_{\langle v \rangle_x\langle w \rangle_x}$ in OCF (a) and CCF (b) around $y_0/h=0.5$ normalized by $(u_\tau h)^2$ for $\mathrm{Re}_\tau\approx900$; Blue solid (red dashed) iso-contours indicate values of $R_{\psi\psi}=+(-)10\textsuperscript{-5}$ to $+(-)6.6\cdot10\textsuperscript{-4}$ with an increment of $5\cdot10\textsuperscript{-5}$. Note that in (a,b) no averaging over the symmetry plane ($\Delta z/h=0$) has been applied in order to show the validity of the statistics.
  (c,d) Spanwise correlation coefficient $\rho_{\psi\psi}$ as defined in equation~\ref{eqn:psi_spanwise_corr} for different $\mathrm{Re}_\tau$ in OCF (c) and CCF (d). Statistics are obtained by averaging over $N=50$ temporally equidistant snapshots in a time interval $\Delta t u_\mathrm{b}/h > 1000$.}
\label{fig:psi_corr}
\end{figure}

In order to extract SSVs from the DNS data, 
we compute 
the streamfunction of the streamwise-averaged crossflow velocity components, 
$\psi_{\langle v \rangle_x\langle w \rangle_x}(y,z,t)$,
\hl{that fulfills the requirement}
% ---------------------------------------------------------------------------- %
\begin{equation}
\begin{pmatrix} 
\partial \psi_{\langle v \rangle_x\langle w \rangle_x} / \partial y  \\ 
\partial \psi_{\langle v \rangle_x\langle w \rangle_x} / \partial z  
\end{pmatrix}
=
\begin{pmatrix} 
-\langle w\rangle_x  \\ 
\langle v\rangle  _x
\end{pmatrix}
        \,.
\end{equation}
% ---------------------------------------------------------------------------- %
%defined as follows:
%% ---------------------------------------------------------------------------- %
%\begin{equation}
%        \psi_{\langle v \rangle_x\langle w \rangle_x}(y,z,t) = \int_{\hat{y}}^y \langle w \rangle_x(\hat{y},z,t) \sd \hat{y} -   \int_{\hat{z}}^{z} \langle v \rangle_x(y,\hat{z},t) \sd \hat{z}
%        \,.
%\end{equation}
%% ---------------------------------------------------------------------------- %
%is taken into account. 
%has been computed. 
%
Figure~\ref{fig:psi_inst} presents instantaneous realizations of the streamfunction for both OCF and CCF 
at arbitrary times.
% ---------------------------------------------------------------------------- %
% ---------------------------------------------------------------------------- %
The iso-contours of the streamfunction in both configurations exhibit vortical structures with a diameter of the order of $h$ which can be associated to SSV. 
In addition, CCF features low-intensity structures spanning the entire channel height $2h$ (figure~\ref{fig:psi_inst}b).
Compared to CCF the structure of the streamfunction in OCF appears to be more coherent with more regularly spanwise-alternating vortices.
In order to compare the structure of the streamfunction quantitatively between OCF and CCF, the two-point correlation of the streamfunction,
% ---------------------------------------------------------------------------- %
\begin{equation}
        R_{\psi\psi}(\Delta y,\Delta z, y_0) = \langle \psi_{\langle v \rangle_x\langle w \rangle_x} (y_0,z) \psi_{\langle v \rangle_x\langle w \rangle_x} (y_0+\Delta y,z+\Delta z) \rangle_{zN},
\end{equation}
% ---------------------------------------------------------------------------- %
where $\langle \cdot \rangle_{zN}$ represents averaging in $z$-direction as well as over $N=50$ snapshots, is presented in figure~\ref{fig:psi_corr}(a,b) for $\mathrm{Re}_\tau=900$.
In addition, the spanwise correlation coefficient of the latter quantity,
% ---------------------------------------------------------------------------- %
\begin{equation}
    \rho_{\psi\psi}(0,\Delta z,y_0)=R_{\psi\psi}(0,\Delta z,y_0)/R_{\psi\psi}(0,0,y_0) 
    \label{eqn:psi_spanwise_corr}
\end{equation}
% ---------------------------------------------------------------------------- %
is displayed in figure~\ref{fig:psi_corr}(c,d) for the range of Reynolds numbers considered here and for 
%the 
a 
reference point located at $y_0/h=0.5$.
% ---------------------------------------------------------------------------- %
% ---------------------------------------------------------------------------- %
Iso-contours of the streamfunction two-point correlation in OCF (figure~\ref{fig:psi_corr}a) clearly reflect the structure of SSVs in terms of alternating streamwise vortices with a radius of $h$ over the full channel width.
For the iso-contours in CCF shown in figure~\ref{fig:psi_corr}(b), on the other hand, the correlation decays at smaller spanwise separation length and the structure is less clear than in OCF.
 This behavior can be even more precisely quantified with the aid of the corresponding spanwise correlation coefficient $\rho_{\psi\psi}$ (figure~\ref{fig:psi_corr}c,d).
While in OCF $\rho_{\psi\psi}$ exhibits a regular alternating patterning at all Reynolds numbers (figure~\ref{fig:psi_corr}c), the spanwise correlation coefficient in CCF hardly features significant secondary minima when the spanwise separation length increases (figure~\ref{fig:psi_corr}d).
Furthermore, the iso-contours of the streamfunction two-point correlation when normalized with a global reference value (figure~\ref{fig:psi_corr}b) indicate that the structures in CCF are less intense than their open channel counterparts (figure~\ref{fig:psi_corr}a).
Both higher correlation values of the streamfunction as well as the spanwise organization of the correlation in OCF can be explained by the free-slip boundary condition, which locks the SSVs in vertical position.
In CCF, on the other hand, these structures are less 
%contained 
constrained 
and 
they 
can propagate (as well as span) across the channel centerline.
Thus, they cancel each other out when averaged and their statistical footprint is less significant than the footprint of corresponding OCF structures.
Note that \citet{Gong2023} reported a similar effect on wall-attached motions, which in OCF are restricted to a region closer to the solid wall than in CCF.

% --- updated surface layer ----------------------------------------------------
\subsection{Updated surface-influenced layer}
\label{ss:layer}
\begin{figure} % LAYER STRUCTURE 
\centering
\includegraphics[]{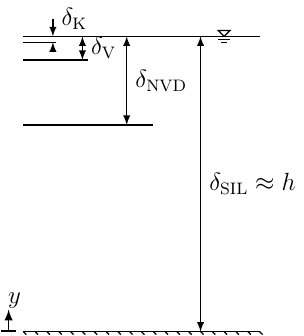}
\caption{Updated surface-influenced layer structure in OCF for high Reynolds numbers. Surface-influenced layer $\delta_{\mathrm{SIL}}\approx h$, normal-velocity-damping layer $\delta_{\mathrm{NVD}}\approx0.3h$, near-surface viscous sublayer $\delta_\mathrm{V} \approx \ell_\mathrm{V}$, Kolmogorov sublayer $\delta_\mathrm{K} \approx 20\ell_\mathrm{K}$.}
\label{fig:layers}
\end{figure}

% ------------------------------------------------------------------------------
The comparison of the streamwise turbulence intensity between OCF and CCF in section~\ref{ss:urms} and SSV in section~\ref{ss:ssv} revealed that the influence of the free-slip boundary condition reaches much further towards the solid wall than 
%initially 
previously 
assumed. 
% here spectra and wall-shear stress? influence all the way down ..
%As a consequence, 
In order to account for this new evidence, 
we separate the ``surface-influenced layer'' from a new ``normal-velocity-damping layer'', thereby expanding the previous three-layer structure to four layers, as sketched in figure~\ref{fig:layers}. 
%we introduce the near-surface layer where the surface-normal turbulence intensity profile in %OCF differs from CCF as normal-velocity-damping layer $\delta_{\mathrm{NVD}}\approx0.3h$, while %expanding the surface-influenced layer to the approximate channel height 
%$\delta_{\mathrm{SIL}}\approx h$.
The thickness of the former layer is found to extend essentially across the entire channel, hence we propose $\delta_{\mathrm{SIL}}\approx h$. The thickness of the latter layer, where the surface-normal turbulence intensity profile in OCF differs from CCF, is quantified by $\delta_{\mathrm{NVD}}\approx0.3h$, similar to previous studies \citep[e.g.][]{Nagaosa1999,Duan2021}. 
On the other hand, the near-surface viscous sublayer, where the surface-normal velocity grows linearly when moving away from the free surface, measures $\delta_\mathrm{V} \approx \ell_\mathrm{V}$, and the Kolmogorov sublayer, where the surface-normal gradient of the mean velocity as well as the intensities of the fluctuations of the surface-parallel vorticity components are damped to zero, is estimated as $\delta_\mathrm{K} \approx 20\ell_\mathrm{K}$.
%A sketch of the updated layer structure is presented in figure~\ref{fig:layers}.
% ---------------------------------------------------------------------------- %
% ---------------------------------------------------------------------------- %
Numerical values for the widths of these four layers as presently defined are given in table~\ref{tab:layers} for the different OCF cases.
% ---------------------------------------------------------------------------- %
\begin{table} % Layer size
\centering
\caption{Surface-influenced layer widths ($\delta_{\mathrm{SIL}}$, $\delta_{\mathrm{NVD}}$, $\delta_\mathrm{V}$, $\delta_\mathrm{K}$).
%and near-surface characteristic length scales ($\ell_\mathrm{V}$, $\ell_\mathrm{K}$).
}
% ----------------------------------------------------------
\begin{tabularx}{1.00\linewidth}{Xccccccc}
\hline\hline
case & $\mathrm{Re}_{\tau}$  & $\delta_{\mathrm{SIL}}/h$ & $\delta_{\mathrm{NVD}}/h$ & $\delta_\mathrm{V}/h$ & $\delta_\mathrm{K}/h$ & 
%$\sfrac{\ell_\mathrm{V}}{h}$ & $\sfrac{\ell_\mathrm{K}}{h}$ & 
$\delta_\mathrm{V}^+$ & $\delta_\mathrm{K}^+$ \\
\hline
O200 & 200  &  1 & 0.3 & 0.0178 & 0.048 & 
%0.0178 & 0.0024 & 
10.7 & 14.2\\
O400 & 399  &  1 & 0.3 & 0.0120 & 0.026 &
%0.0120 & 0.0013 & 
14.3 & 15.7\\
O600 & 596  &  1 & 0.3 & 0.0095 & 0.018 &
%0.0095 & 0.0009 & 
17.1 & 16.6\\
O900 & 895  &  1 & 0.3 & 0.0076 & 0.014 &
%0.0076 & 0.0007 & 
20.3 & 17.6\\
\hline\hline
\end{tabularx}
% ----------------------------------------------------------
\label{tab:layers}
\end{table}
% ----------------------------------------------------------
Note that $\delta_\mathrm{K} > \delta_\mathrm{V}$ in the present range of Reynolds numbers, while  
%For the proposed layer widths 
Reynolds numbers $\mathrm{Re}_\mathrm{b}\gtrsim 160000$ are necessary for $\delta_\mathrm{K} < \delta_\mathrm{V}$. 
Hence the relative widths of $\delta_\mathrm{V}$ and $\delta_\mathrm{K}$ sketched in figure~\ref{fig:layers} apply to high-Reynolds number OCF.

The Kolmogorov layer width, which varies only little when normalized in wall units for the range of Reynolds numbers considered ($14.2 \le \delta_\mathrm{K}^+ \le 17.6$), is of the same order of magnitude as the wall-normal extent of the viscous sublayer adjacent to the no-slip wall.
This highlights a numerical resolution requirement for the free-slip boundary similar to the one near the no-slip wall. 
DNSs of OCF often use one-sided grid refinement only, which leads to acceptable one-point turbulence statistics near the solid wall.
However, the non-linear interactions between the small scales, which are 
%settled 
located 
in the near-surface region, and larger scales of motions might not be captured correctly without additional grid refinement towards the free-slip boundary. 
%The observed scaling clearly shows a necessity of the fine grid resolutions near the free-slip plane which are employed in this study, to truthfully represent the first-order statistics already.
%As the major part of the inter-component energy transfer takes place within a thin layer scaling with $\ell_\mathrm{V}$, this is another confirmation that the DNS of open channel flow requires grid refinement towards the free-slip boundary.
%The present two-sided grid refinement is evaluated in appendix~\ref{app:gridrefinement}.
%
% --- Conclusion ---------------------------------------------------------------
\section{Conclusion}
% ------------------------------------------------------------------------------
In turbulent open channel flow the free surface is typically modelled as a free-slip boundary when it can be expected to undergo only small deformations. 
As a consequence the statistics 
generally 
differ from the counterpart in closed channel flow. 
For a long time the differences in these two configurations were believed to be restricted to the vicinity of the free surface.
The most prominent effect of the free-slip boundary condition is on the structure of the Reynolds stress tensor, which has been widely discussed in the literature~\citep{Swean1991,Handler1993,Pan1995,Nagaosa1997,Nagaosa1999,Handler1999,Calmet2003,Nagaosa2003,Nagaosa2012}.
In this work, the scaling of turbulence statistics in the vicinity of a free-slip boundary in open channel flow has been analyzed with the aid of data from DNS in computational domains large enough to capture the largest turbulence scales while providing sufficient grid refinement near 
%the no-slip and 
the free-slip boundary.
For $\mathrm{Re}_\tau$ up to 900, the triple surface layer proposed by CM03 has been observed for the spanwise and wall-normal Reynolds stress components as well as for the vorticity fluctuation intensities. 
In accordance with CM03 we found 
the channel height 
$h$, 
the viscous length 
$\ell_\mathrm{V}$, and 
the Kolmogorov length 
$\ell_\mathrm{K}$ to be the relevant length scales for the near-surface layers.
Unlike CM03, however, the sublayer where the wall-normal velocity decreases linearly towards the surface is found to scale with the near-surface viscous length scale $\ell_\mathrm{V}$ (instead of $\ell_\mathrm{K}$), and the thin near-surface vorticity layer is found to scale with the Kolmogorov length scale $\ell_\mathrm{K}$ (instead of $\ell_\mathrm{V}$).
%Moreover, as we reported in \cite{Bauer2023}, the streamwise Reynolds stress component
% % ---------------------------------------------------------------------------- %
% The vorticity intensity was found to scale in wall units times $Re_\tau^{1/2}$, and the terms in the TKE budget, as well as the pressure-rate-of-strain tensor, were found to scale in wall units times $Re_\tau$, as suggested by CM03 and \citet{Hoyas2008}.\par
% % ---------------------------------------------------------------------------- %
% The relevant length scale for the TKE budget terms in the vicinity of the free surface is either the Kolmogorov scale $\ell_\mathrm{K}$ (for the turbulent dissipation term) or the surface-near viscous scale $\ell_\mathrm{V}$ (viscous, turbulent, and pressure diffusion).
% The fact that the transport terms scale with $\ell_\mathrm{V}$ (as the Reynolds stresses) and the dissipation term scales with $\ell_\mathrm{K}$ (as the vorticity fluctuation intensity) reflects the contribution of different turbulent coherent structures to the corresponding terms.
% The transport of TKE in the vicinity of the free surface seems to be related to large ``streaky'' structures in the velocity field, whereas its dissipation is organized by rather small-scale vortical motions.\par
% ----------------------------------------------------------------------------
Regarding their intensity, spanwise and wall-normal velocity fluctuations 
do 
scale in wall units, while the streamwise component tends to increase under that scaling when increasing the Reynolds number.
This failure of wall-scaling is well known for closed channel flow and it is related to VLSMs that contribute to the streamwise Reynolds stress component for higher Reynolds numbers.
For open channel flows this effect appears already at lower Reynolds numbers 
%
%($Re_\tau{\geq}400$) 
%
and leads---unlike for closed channel flows---to a collapse of the streamwise turbulent intensity profiles normalized with the bulk velocity 
in the region 
between $y/h \approx 0.3$ and $y/h = 1$, which is already visible for $\mathrm{Re}_\tau{\ge}400$.
% ---------------------------------------------------------------------------- %
Since the footprint of the enhanced VLSMs can be found in energy spectra very close to the solid wall, the so-called surface-influenced layer $\delta_{\mathrm{SIL}}$, described by CM03 and others, has to be extended essentially all the way to the solid boundary for OCF.
As a consequence, turbulence in the vicinity of a free-slip boundary in OCF has effectively to be characterized by four layers: 
The Kolmogorov sublayer $\delta_\mathrm{K}\approx 20\mathrm{Re}_\mathrm{b}^{-3/4}h$---where surface-parallel vorticities are damped to zero---, 
the viscous sublayer $\delta_\mathrm{V}\approx\mathrm{Re}_\mathrm{b}^{-1/2}h$---where the surface-normal velocity component decreases linearly to zero---,  
the normal velocity damping layer $\delta_{\mathrm{NVD}}\approx0.3h$---where the surface-normal velocity profile deviates from the CCF profile---
and the surface-influenced layer 
%being the 
spanning the entire 
channel height $\delta_{\mathrm{SIL}}=h$---where the streamwise turbulence intensity profile deviates from the CCF profile.\par
% ---------------------------------------------------------------------------- %
Note that the above mentioned effect on the streamwise velocity profile cannot be faithfully observed in numerical simulations of turbulent open channel flow either at low Reynolds number---where VLSM are not present---
%or not energetic---
or in small computational domains---where VLSM are artificially suppressed.
% ============================================================================ %
%
%
%
\backmatter

%\bmhead{Supplementary information}
%
%If your article has accompanying supplementary file/s please state so here. 
%
%Authors reporting data from electrophoretic gels and blots should supply the full unprocessed scans for key as part of their Supplementary information. This may be requested by the editorial team/s if it is missing.
%
%Please refer to Journal-level guidance for any specific requirements.
%
\bmhead{Acknowledgements}
Many fruitful discussions with Genta Kawahara throughout this work are gratefully acknowledged. 
The simulations were carried out at the HPC clusters UC2 and CARA. Thus, 
the authors gratefully acknowledge the scientific support and HPC resources provided by the SCC Karlstuhe, the German Aerospace Center (DLR), and the state of Baden-Wuerttemberg through bwHPC. 
The HPC system CARA is partially funded by "Saxon State Ministry for Economic Affairs, Labour and Transport" and "Federal Ministry for Economic Affairs and Climate Action".
%
%The authors acknowledge funding by DFG through grant UH~242/3-1.
%

 \section*{Declarations}
% 
%Some journals require declarations to be submitted in a standardised format. Please check the Instructions for Authors of the journal to which you are submitting to see if you need to complete this section. If yes, your manuscript must contain the following sections under the heading `Declarations':
%
\begin{itemize}
\item Funding: The authors acknowledge funding by DFG through grant UH~242/3-1.
%\item Conflict of interest/Competing interests (check journal-specific guidelines for which heading to use)
%\item Ethics approval and consent to participate
%\item Consent for publication
%\item Data availability 
%\item Materials availability
%\item Code availability 
%\item Author contribution
\end{itemize}

%\noindent
%If any of the sections are not relevant to your manuscript, please include the heading and write `Not applicable' for that section. 

%%===================================================%%
%% For presentation purpose, we have included        %%
%% \bigskip command. Please ignore this.             %%
%%===================================================%%
%%\bigskip
%\begin{flushleft}%
%Editorial Policies for:

%\bigskip\noindent
%Springer journals and proceedings: \url{https://www.springer.com/gp/editorial-policies}

%\bigskip\noindent
%Nature Portfolio journals: \url{https://www.nature.com/nature-research/editorial-policies}

%\bigskip\noindent
%\textit{Scientific Reports}: \url{https://www.nature.com/srep/journal-policies/editorial-policies}

%\bigskip\noindent
%BMC journals: \url{https://www.biomedcentral.com/getpublished/editorial-policies}
%\end{flushleft}

% === Appendix =============================================================== %
%\appendix
\begin{appendices}
% ---------------------------------------------------------------------------- %
%
% ---------------------------------------------------------------------------- %
\section{Grid refinement towards the boundaries}
\label{app:gridrefinement}
\begin{figure}% DELTAY/ETA
\centering
\includegraphics[]{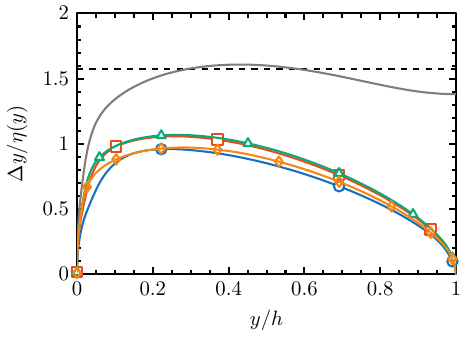}
\caption{Ratio of the vertical spacing $\Delta y$ and the local Kolmogorov scale $\eta(y) = (\nu^3/\varepsilon(y))^{1/4}$ . \symba, O200; \symbb, O400; \symbc, O600; \symbd, O900. The gray line shows closed channel reference data  \citep{DelAlamo2003} at  $\mathrm{Re}_{\tau}=550$. The dashed line indicates a reference value of $\pi/2$, commonly employed in DNS of homogeneous isotropic turbulence and initially stated by \citet{Jimenez1993}.}
\label{fig:feta}
\end{figure}

\begin{figure} % RS ASYMPTOTE
\centering
\includegraphics[width=\linewidth]{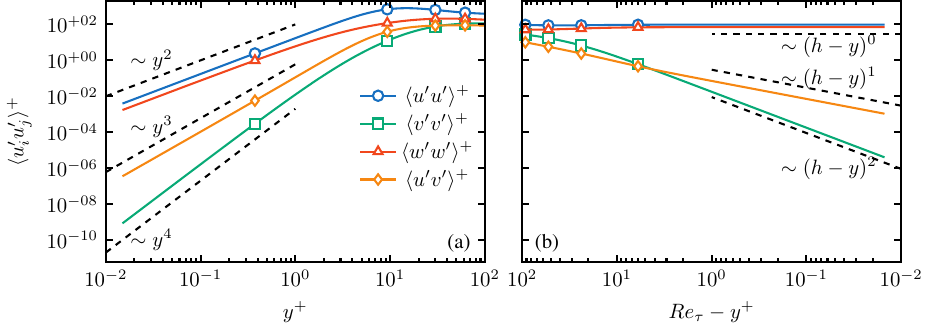}
\caption{Asymptotic behavior of Reynolds stress components for O900. \symbx, $\langle u'u' \rangle^+$; \symby, $\langle v'v' \rangle^+$; \symbz, $\langle w'w' \rangle^+$; \symbxy, $\langle u'v' \rangle^+$. (a): Near-wall region, (b): near-surface region.}
\label{fig:reyasy}
\end{figure}

% ---------------------------------------------------------------------------- %
As mentioned above, DNSs of turbulent open channel flow often lack grid refinement towards the free-slip boundary.
However, the thin velocity and vorticity damping layer near the free-slip boundary, as well as the underlying small-scale coherent motions, make a grid refinement towards the free-slip surface desirable.
Figure~\ref{fig:feta} shows that the vertical spacing $\Delta y$ is kept below the value of the Kolmogorov length scale $\eta$ throughout most of the flow domain.
% ---------------------------------------------------------------------------- %
% ---------------------------------------------------------------------------- %
In order to evaluate whether the boundary conditions near the no-slip and the free-slip boundary are fulfilled, the wall-normal asymptotics of the Reynolds stress components---which can be obtained from a Taylor series expansion and making use of the boundary conditions as well as of the continuity equation---are taken into account.
Figure \ref{fig:reyasy} shows that the expected asymptotic behavior is well established both near the solid wall and near the free-slip surface.
% ---------------------------------------------------------------------------- %
% ---------------------------------------------------------------------------- %
Therefore, the sufficiency of the near-boundary grid resolution is validated in this context.\par
% ---------------------------------------------------------------------------- %
%
%
% === Very-large-scale motions =============================================== %
\section{Very-large-scale motions}
\label{app:vlsm}
\begin{figure} % RUU3D
\centering
\includegraphics[]{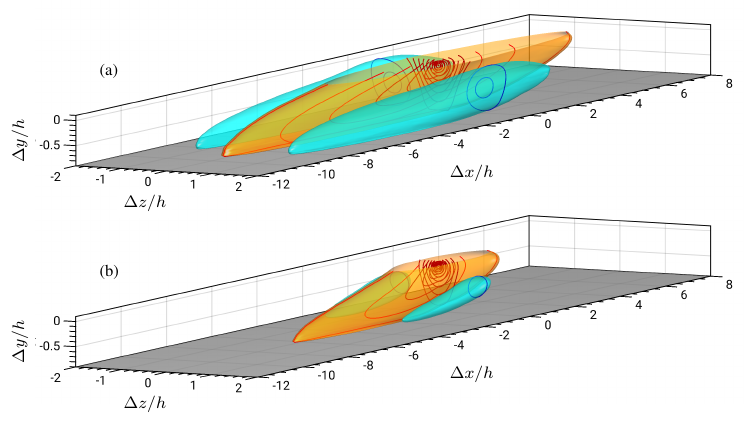}
\caption{Iso-surfaces and iso-contours of the three-dimensional two-point velocity correlation of the streamwise velocity component $\rho_{uu}$ as a function of the streamwise, spanwise and wall-normal separation lengths. Reference point of the correlation lies at $y_0/R=0.9$. (a) O900; (b) C900. The orange (cyan) iso-surfaces exhibit values of $+(-)0.1$. Iso-contours are shown for the $xy$-plane at $z/h=0$ and the $yz$-plane at 
%$z/h=0$ 
$x/h=0$ 
with values ranging from $+(-)0.1$ to $+(-)1$, increment of $0.1$.}
\label{fig:ruu3d}
\end{figure}

\begin{figure} % 2D PMS different Re
\centering
\includegraphics[]{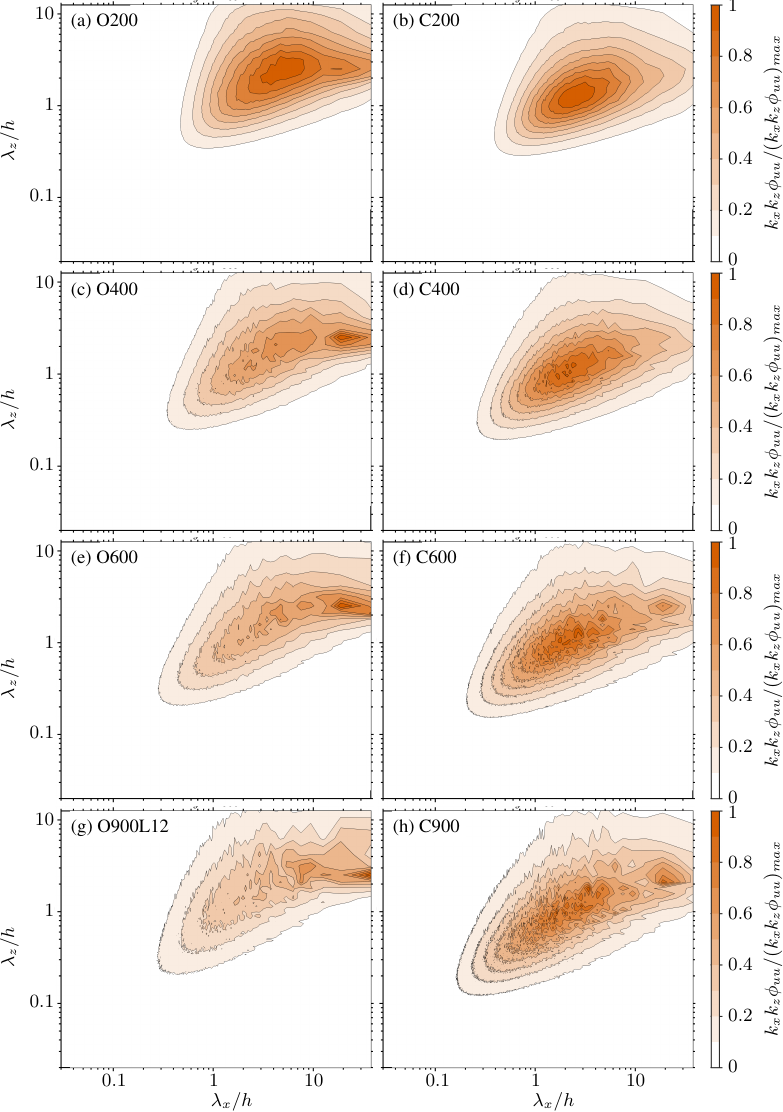}
\caption{Contour lines of two-dimensional pre-multiplied energy spectra of streamwise velocity fluctuations at $y/h\approx0.95$ as a function of the stream- and spanwise wavelengths $\kappa_x\kappa_z\phi^+_{uu}(\lambda_x,\lambda_z)$. Left column, open channel; right column, closed channel. (a,b) $\mathrm{Re}_\tau\approx200$; (c,d) $\mathrm{Re}_\tau\approx400$; (e,f) $\mathrm{Re}_\tau\approx 600$; (g,h) $\mathrm{Re}_\tau\approx 900$.}
\label{fig:pms2d}
\end{figure}

Statistically, VLSMs can be extracted from three-dimensional two-point velocity correlations of the streamwise velocity component with a reference point at $y/h=0.9$ for O900 and C900, as displayed in figure~\ref{fig:ruu3d}.
% ---------------------------------------------------------------------------- %
% ---------------------------------------------------------------------------- %
In agreement with \citet{Duan2020,Duan2021}, the iso-surface of the streamwise velocity correlation ($\rho_{uu}=0.1$) indicates, that large-scale coherent structures in OCF are longer and wider than their closed channel counterpart.
% ---------------------------------------------------------------------------- %
More detailed information is obtained through the analysis of pre-multiplied energy spectra of the different Reynolds number cases for OCF and CCF.
The analysis of pre-multiplied energy spectra to estimate the spatial extent of coherent structures
has been performed in various CCF studies~\citep{Jimenez1998,DelAlamo2003,Abe2004,Monty2007,Monty2009} as well as recent OCF studies by \citet{Duan2020,Duan2021}.
The location of the peak in a pre-multiplied energy spectrum corresponds to the wavelength of the most energetic motions~\citep{Perry1986}.
In figure \ref{fig:pms2d} two-dimensional pre-multiplied spectra of the streamwise velocity fluctuation below the free-slip boundary ($y/h=0.95$) are compared to the corresponding closed channel data.
% ---------------------------------------------------------------------------- %
% ---------------------------------------------------------------------------- %
As mentioned by \citet{Duan2020,Duan2021}, the location of the outer spectral peak indicates that large-scale coherent structures in OCF are longer and wider than their closed channel counterpart.
In general, energetic coherent structures in the vicinity of the free-slip boundary in OCF appear at spanwise wavelengths that are about a factor of two larger than those related to coherent structures in CCF.
Furthermore, in agreement with \citet{Pinelli2022}, the outer spectral peak related to VLSM in OCF is visible for Reynolds numbers as low as 400 (figure~\ref{fig:pms2d}c), whereas in comparable CCF configurations the outer spectral peak occurs for $\mathrm{Re}_\tau \gtrsim 600$ (figure~\ref{fig:pms2d}f).
In \citet{Duan2021} it was noted that VLSMs contribute more to the wall-shear stress in OCF than in CCF, while LSMs contribute less to wall-shear stress in OCF than in CCF.
The two-dimensional energy spectra in figure~\ref{fig:pms2d} support this finding, since in CCF the spectral energy peak related to large-scale motions visible at length scales $\lambda_x\approx 2h,\lambda_z\approx h$ appears to be 
%larger 
of higher amplitude 
than the one related to VLSM (figure~\ref{fig:pms2d}f,h).
In OCF, on the contrary, the VLSM peak clearly dominates the spectra (figure~\ref{fig:pms2d}c,e,g).\par

\end{appendices}
% === Bibliograpghy ========================================================== %
\clearpage
\bibliography{library}

%% BioMed_Central_Bib_Style_v1.01

\begin{thebibliography}{42}
% BibTex style file: bmc-mathphys.bst (version 2.1), 2014-07-24
\ifx \bisbn   \undefined \def \bisbn  #1{ISBN #1}\fi
\ifx \binits  \undefined \def \binits#1{#1}\fi
\ifx \bauthor  \undefined \def \bauthor#1{#1}\fi
\ifx \batitle  \undefined \def \batitle#1{#1}\fi
\ifx \bjtitle  \undefined \def \bjtitle#1{#1}\fi
\ifx \bvolume  \undefined \def \bvolume#1{\textbf{#1}}\fi
\ifx \byear  \undefined \def \byear#1{#1}\fi
\ifx \bissue  \undefined \def \bissue#1{#1}\fi
\ifx \bfpage  \undefined \def \bfpage#1{#1}\fi
\ifx \blpage  \undefined \def \blpage #1{#1}\fi
\ifx \burl  \undefined \def \burl#1{\textsf{#1}}\fi
\ifx \doiurl  \undefined \def \doiurl#1{\url{https://doi.org/#1}}\fi
\ifx \betal  \undefined \def \betal{\textit{et al.}}\fi
\ifx \binstitute  \undefined \def \binstitute#1{#1}\fi
\ifx \binstitutionaled  \undefined \def \binstitutionaled#1{#1}\fi
\ifx \bctitle  \undefined \def \bctitle#1{#1}\fi
\ifx \beditor  \undefined \def \beditor#1{#1}\fi
\ifx \bpublisher  \undefined \def \bpublisher#1{#1}\fi
\ifx \bbtitle  \undefined \def \bbtitle#1{#1}\fi
\ifx \bedition  \undefined \def \bedition#1{#1}\fi
\ifx \bseriesno  \undefined \def \bseriesno#1{#1}\fi
\ifx \blocation  \undefined \def \blocation#1{#1}\fi
\ifx \bsertitle  \undefined \def \bsertitle#1{#1}\fi
\ifx \bsnm \undefined \def \bsnm#1{#1}\fi
\ifx \bsuffix \undefined \def \bsuffix#1{#1}\fi
\ifx \bparticle \undefined \def \bparticle#1{#1}\fi
\ifx \barticle \undefined \def \barticle#1{#1}\fi
\bibcommenthead
\ifx \bconfdate \undefined \def \bconfdate #1{#1}\fi
\ifx \botherref \undefined \def \botherref #1{#1}\fi
\ifx \url \undefined \def \url#1{\textsf{#1}}\fi
\ifx \bchapter \undefined \def \bchapter#1{#1}\fi
\ifx \bbook \undefined \def \bbook#1{#1}\fi
\ifx \bcomment \undefined \def \bcomment#1{#1}\fi
\ifx \oauthor \undefined \def \oauthor#1{#1}\fi
\ifx \citeauthoryear \undefined \def \citeauthoryear#1{#1}\fi
\ifx \endbibitem  \undefined \def \endbibitem {}\fi
\ifx \bconflocation  \undefined \def \bconflocation#1{#1}\fi
\ifx \arxivurl  \undefined \def \arxivurl#1{\textsf{#1}}\fi
\csname PreBibitemsHook\endcsname

%%% 1
\bibitem[\protect\citeauthoryear{Abe et~al.}{2004}]{Abe2004}
\begin{barticle}
\bauthor{\bsnm{Abe}, \binits{H.}},
\bauthor{\bsnm{Kawamura}, \binits{H.}},
\bauthor{\bsnm{Choi}, \binits{H.}}:
\batitle{Very large-scale structures and their effects on the wall shear-stress
  fluctuations in a turbulent channel flow up to
  {{Re}}{\textsubscript{{$\tau$}}}=640}.
\bjtitle{Journal of Fluids Engineering}
\bvolume{126}(\bissue{5}),
\bfpage{835}--\blpage{843}
(\byear{2004})
\doiurl{10.1115/1.1789528}
\end{barticle}
\endbibitem

%%% 2
\bibitem[\protect\citeauthoryear{Antonia et~al.}{1996}]{Antonia1996}
\begin{barticle}
\bauthor{\bsnm{Antonia}, \binits{R.A.}},
\bauthor{\bsnm{Rajagopalan}, \binits{S.}},
\bauthor{\bsnm{Zhu}, \binits{Y.}}:
\batitle{Scaling of mean square vorticity in turbulent flows}.
\bjtitle{Experiments in Fluids}
\bvolume{20}(\bissue{5}),
\bfpage{393}--\blpage{394}
(\byear{1996})
\doiurl{10.1007/BF00191021}
\end{barticle}
\endbibitem

%%% 3
\bibitem[\protect\citeauthoryear{Bauer}{2015}]{Bauer2015}
\begin{botherref}
\oauthor{\bsnm{Bauer}, \binits{C.}}:
Direct numerical simulation of turbulent open channel flow.
Master's thesis,
Karlsruher Institut f{\"u}r Technologie
(2015)
\end{botherref}
\endbibitem

%%% 4
\bibitem[\protect\citeauthoryear{Brumley and Jirka}{1988}]{Brumley1988}
\begin{barticle}
\bauthor{\bsnm{Brumley}, \binits{B.H.}},
\bauthor{\bsnm{Jirka}, \binits{G.H.}}:
\batitle{Air-water transfer of slightly soluble gases: Turbulence, interfacial
  processes and conceptual models}.
\bjtitle{Physicochemical hydrodynamics}
\bvolume{10}(\bissue{3}),
\bfpage{295}--\blpage{319}
(\byear{1988})
\end{barticle}
\endbibitem

%%% 5
\bibitem[\protect\citeauthoryear{Borue et~al.}{1995}]{Borue1995}
\begin{barticle}
\bauthor{\bsnm{Borue}, \binits{V.}},
\bauthor{\bsnm{Orszag}, \binits{S.A.}},
\bauthor{\bsnm{Staroselsky}, \binits{I.}}:
\batitle{Interaction of surface waves with turbulence: Direct numerical
  simulations of turbulent open-channel flow}.
\bjtitle{Journal of Fluid Mechanics}
\bvolume{286},
\bfpage{1}--\blpage{23}
(\byear{1995})
\end{barticle}
\endbibitem

%%% 6
\bibitem[\protect\citeauthoryear{Bauer et~al.}{2023}]{BauerData2023}
\begin{botherref}
\oauthor{\bsnm{Bauer}, \binits{C.}},
\oauthor{\bsnm{Sakai}, \binits{Y.}},
\oauthor{\bsnm{Uhlmann}, \binits{M.}}:
Data Underlying the Publication: {{Direct}} Numerical Simulation of Turbulent
  Open Channel Flow,
  {{https://doi.org/10.4121/88678f02-2a34-4452-8534-6361fc34d06b}}.
4TU.ResearchData
(2023).
\doiurl{10.4121/88678f02-2a34-4452-8534-6361fc34d06b}
\end{botherref}
\endbibitem

%%% 7
\bibitem[\protect\citeauthoryear{Bauer et~al.}{2024}]{Bauer2024b}
\begin{bchapter}
\bauthor{\bsnm{Bauer}, \binits{C.}},
\bauthor{\bsnm{Sakai}, \binits{Y.}},
\bauthor{\bsnm{Uhlmann}, \binits{M.}}:
\bctitle{Direct numerical simulation of turbulent open channel flow:
  {{Streamwise}} turbulence intensity scaling and its relation to large-scale
  coherent motions}.
In: \beditor{\bsnm{{\"O}rl{\"u}}, \binits{R.}},
\beditor{\bsnm{Talamelli}, \binits{A.}},
\beditor{\bsnm{Peinke}, \binits{J.}},
\beditor{\bsnm{Oberlack}, \binits{M.}} (eds.)
\bbtitle{Progress in Turbulence {{X}}},
pp. \bfpage{311}--\blpage{317}.
\bpublisher{Springer},
\blocation{Cham}
(\byear{2024}).
\doiurl{10.1007/978-3-031-55924-2_42}
\end{bchapter}
\endbibitem

%%% 8
\bibitem[\protect\citeauthoryear{Calmet and Magnaudet}{2003}]{Calmet2003}
\begin{barticle}
\bauthor{\bsnm{Calmet}, \binits{I.}},
\bauthor{\bsnm{Magnaudet}, \binits{J.}}:
\batitle{Statistical structure of high-{{Reynolds-number}} turbulence close to
  the free surface of an open-channel flow}.
\bjtitle{Journal of Fluid Mechanics}
\bvolume{474},
\bfpage{355}--\blpage{378}
(\byear{2003})
\doiurl{10.1017/S0022112002002793}
\end{barticle}
\endbibitem

%%% 9
\bibitem[\protect\citeauthoryear{Duan et~al.}{2020}]{Duan2020}
\begin{barticle}
\bauthor{\bsnm{Duan}, \binits{Y.}},
\bauthor{\bsnm{Chen}, \binits{Q.}},
\bauthor{\bsnm{Li}, \binits{D.}},
\bauthor{\bsnm{Zhong}, \binits{Q.}}:
\batitle{Contributions of very large-scale motions to turbulence statistics in
  open channel flows}.
\bjtitle{Journal of Fluid Mechanics}
\bvolume{892},
\bfpage{3}
(\byear{2020})
\doiurl{10.1017/jfm.2020.174}
\end{barticle}
\endbibitem

%%% 10
\bibitem[\protect\citeauthoryear{{del {\'A}lamo} and
  Jim{\'e}nez}{2003}]{DelAlamo2003}
\begin{barticle}
\bauthor{\bsnm{{del {\'A}lamo}}, \binits{J.C.}},
\bauthor{\bsnm{Jim{\'e}nez}, \binits{J.}}:
\batitle{Spectra of the very large anisotropic scales in turbulent channels}.
\bjtitle{Physics of Fluids}
\bvolume{15}(\bissue{6}),
\bfpage{41}--\blpage{44}
(\byear{2003})
\doiurl{10.1063/1.1570830}
\end{barticle}
\endbibitem

%%% 11
\bibitem[\protect\citeauthoryear{{del {\'A}lamo} et~al.}{2004}]{DelAlamo2004}
\begin{barticle}
\bauthor{\bsnm{{del {\'A}lamo}}, \binits{J.C.}},
\bauthor{\bsnm{Jim{\'e}nez}, \binits{J.}},
\bauthor{\bsnm{Zandonade}, \binits{P.}},
\bauthor{\bsnm{Moser}, \binits{R.D.}}:
\batitle{Scaling of the energy spectra of turbulent channels}.
\bjtitle{Journal of Fluid Mechanics}
\bvolume{500},
\bfpage{135}--\blpage{144}
(\byear{2004})
\doiurl{10.1017/S002211200300733X}
\end{barticle}
\endbibitem

%%% 12
\bibitem[\protect\citeauthoryear{Duan et~al.}{2021}]{Duan2021}
\begin{barticle}
\bauthor{\bsnm{Duan}, \binits{Y.}},
\bauthor{\bsnm{Zhong}, \binits{Q.}},
\bauthor{\bsnm{Wang}, \binits{G.}},
\bauthor{\bsnm{Zhang}, \binits{P.}},
\bauthor{\bsnm{Li}, \binits{D.}}:
\batitle{Contributions of different scales of turbulent motions to the mean
  wall-shear stress in open channel flows at low-to-moderate {{Reynolds}}
  numbers}.
\bjtitle{Journal of Fluid Mechanics}
\bvolume{918},
\bfpage{40}
(\byear{2021})
\doiurl{10.1017/jfm.2021.236}
\end{barticle}
\endbibitem

%%% 13
\bibitem[\protect\citeauthoryear{Gong et~al.}{2023}]{Gong2023}
\begin{barticle}
\bauthor{\bsnm{Gong}, \binits{Z.}},
\bauthor{\bsnm{Duan}, \binits{Y.}},
\bauthor{\bsnm{Chen}, \binits{X.}},
\bauthor{\bsnm{Li}, \binits{D.}},
\bauthor{\bsnm{Fu}, \binits{X.}}:
\batitle{Statistical behavior of wall-attached motions in open- and
  closed-channel flows via direct numerical simulation}.
\bjtitle{Physics of Fluids}
\bvolume{35}(\bissue{4}),
\bfpage{045138}
(\byear{2023})
\doiurl{10.1063/5.0144392}
\end{barticle}
\endbibitem

%%% 14
\bibitem[\protect\citeauthoryear{Hunt and Graham}{1978}]{Hunt1978}
\begin{barticle}
\bauthor{\bsnm{Hunt}, \binits{J.C.R.}},
\bauthor{\bsnm{Graham}, \binits{J.M.R.}}:
\batitle{Free-stream turbulence near plane boundaries}.
\bjtitle{Journal of Fluid Mechanics}
\bvolume{84}(\bissue{02}),
\bfpage{209}--\blpage{235}
(\byear{1978})
\doiurl{10.1017/S0022112078000130}
\end{barticle}
\endbibitem

%%% 15
\bibitem[\protect\citeauthoryear{Hoyas and Jim{\'e}nez}{2006}]{Hoyas2006}
\begin{barticle}
\bauthor{\bsnm{Hoyas}, \binits{S.}},
\bauthor{\bsnm{Jim{\'e}nez}, \binits{J.}}:
\batitle{Scaling of the velocity fluctuations in turbulent channels up to
  {{Re}}{\textsubscript{{$\tau$}}}=2003}.
\bjtitle{Physics of Fluids}
\bvolume{18}(\bissue{1}),
\bfpage{11702}
(\byear{2006})
\doiurl{10.1063/1.2162185}
\end{barticle}
\endbibitem

%%% 16
\bibitem[\protect\citeauthoryear{Handler et~al.}{1999}]{Handler1999}
\begin{barticle}
\bauthor{\bsnm{Handler}, \binits{R.A.}},
\bauthor{\bsnm{Saylor}, \binits{J.R.}},
\bauthor{\bsnm{Leighton}, \binits{R.I.}},
\bauthor{\bsnm{Rovelstad}, \binits{A.L.}}:
\batitle{Transport of a passive scalar at a shear-free boundary in fully
  developed turbulent open channel flow}.
\bjtitle{Physics of Fluids (1994-present)}
\bvolume{11}(\bissue{9}),
\bfpage{2607}--\blpage{2625}
(\byear{1999})
\doiurl{10.1063/1.870123}
\end{barticle}
\endbibitem

%%% 17
\bibitem[\protect\citeauthoryear{Handler et~al.}{1993}]{Handler1993}
\begin{barticle}
\bauthor{\bsnm{Handler}, \binits{R.A.}},
\bauthor{\bsnm{Swean}, \binits{T.F.}},
\bauthor{\bsnm{Leighton}, \binits{R.I.}},
\bauthor{\bsnm{Swearingen}, \binits{J.D.}}:
\batitle{Length scales and the energy balance for turbulence near a free
  surface}.
\bjtitle{AIAA Journal}
\bvolume{31}(\bissue{11}),
\bfpage{1998}--\blpage{2007}
(\byear{1993})
\doiurl{10.2514/3.11883}
\end{barticle}
\endbibitem

%%% 18
\bibitem[\protect\citeauthoryear{Hunt}{1984}]{Hunt1984}
\begin{bbook}
\bauthor{\bsnm{Hunt}, \binits{J.C.R.}}:
In: \beditor{\bsnm{Brutsaert}, \binits{W.}},
\beditor{\bsnm{Jirka}, \binits{G.H.}} (eds.)
\bbtitle{Turbulence Structure and Turbulent Diffusion Near Gas-Liquid
  Interfaces},
pp. \bfpage{67}--\blpage{82}.
\bpublisher{Springer},
\blocation{Dordrecht}
(\byear{1984}).
\doiurl{10.1007/978-94-017-1660-4_7}
\end{bbook}
\endbibitem

%%% 19
\bibitem[\protect\citeauthoryear{Jim{\'e}nez}{1998}]{Jimenez1998}
\begin{botherref}
\oauthor{\bsnm{Jim{\'e}nez}, \binits{J.}}:
The largest scales of turbulent wall flows.
Technical report,
CTR Annual Research Briefs
(1998)
\end{botherref}
\endbibitem

%%% 20
\bibitem[\protect\citeauthoryear{Jim{\'e}nez and Pinelli}{1999}]{Jimenez1999}
\begin{barticle}
\bauthor{\bsnm{Jim{\'e}nez}, \binits{J.}},
\bauthor{\bsnm{Pinelli}, \binits{A.}}:
\batitle{The autonomous cycle of near-wall turbulence}.
\bjtitle{Journal of Fluid Mechanics}
\bvolume{389},
\bfpage{335}--\blpage{359}
(\byear{1999})
\doiurl{10.1017/S0022112099005066}
\end{barticle}
\endbibitem

%%% 21
\bibitem[\protect\citeauthoryear{Jim{\'e}nez et~al.}{1993}]{Jimenez1993}
\begin{barticle}
\bauthor{\bsnm{Jim{\'e}nez}, \binits{J.}},
\bauthor{\bsnm{Wray}, \binits{A.A.}},
\bauthor{\bsnm{Saffman}, \binits{P.G.}},
\bauthor{\bsnm{Rogallo}, \binits{R.S.}}:
\batitle{The structure of intense vorticity in isotropic turbulence}.
\bjtitle{Journal of Fluid Mechanics}
\bvolume{255},
\bfpage{65}--\blpage{90}
(\byear{1993})
\doiurl{10.1017/S0022112093002393}
\end{barticle}
\endbibitem

%%% 22
\bibitem[\protect\citeauthoryear{Kim et~al.}{1987}]{Kim1987}
\begin{barticle}
\bauthor{\bsnm{Kim}, \binits{J.}},
\bauthor{\bsnm{Moin}, \binits{P.}},
\bauthor{\bsnm{Moser}, \binits{R.}}:
\batitle{Turbulence statistics in fully developed channel flow at low
  {{Reynolds}} number}.
\bjtitle{Journal of Fluid Mechanics}
\bvolume{177},
\bfpage{133}--\blpage{166}
(\byear{1987})
\doiurl{10.1017/S0022112087000892}
\end{barticle}
\endbibitem

%%% 23
\bibitem[\protect\citeauthoryear{Komori et~al.}{1993}]{Komori1993}
\begin{barticle}
\bauthor{\bsnm{Komori}, \binits{S.}},
\bauthor{\bsnm{Nagaosa}, \binits{R.}},
\bauthor{\bsnm{Murakami}, \binits{Y.}},
\bauthor{\bsnm{Chiba}, \binits{S.}},
\bauthor{\bsnm{Ishii}, \binits{K.}},
\bauthor{\bsnm{Kuwahara}, \binits{K.}}:
\batitle{Direct numerical simulation of three-dimensional open-channel flow
  with zero-shear gas--liquid interface}.
\bjtitle{Physics of Fluids A: Fluid Dynamics}
\bvolume{5}(\bissue{1}),
\bfpage{115}--\blpage{125}
(\byear{1993})
\doiurl{10.1063/1.858797}
\end{barticle}
\endbibitem

%%% 24
\bibitem[\protect\citeauthoryear{Lee and Moser}{2015}]{Lee2015}
\begin{barticle}
\bauthor{\bsnm{Lee}, \binits{M.}},
\bauthor{\bsnm{Moser}, \binits{R.D.}}:
\batitle{Direct numerical simulation of turbulent channel flow up to
  {{Re}}{\textsubscript{{$\tau$}}}=5200}.
\bjtitle{Journal of Fluid Mechanics}
\bvolume{774},
\bfpage{395}--\blpage{415}
(\byear{2015})
\doiurl{10.1017/jfm.2015.268}
\end{barticle}
\endbibitem

%%% 25
\bibitem[\protect\citeauthoryear{Leighton et~al.}{1991}]{Leighton1991}
\begin{bchapter}
\bauthor{\bsnm{Leighton}, \binits{R.I.}},
\bauthor{\bsnm{Swean~Jr}, \binits{T.F.}},
\bauthor{\bsnm{Handler}, \binits{R.A.}},
\bauthor{\bsnm{Swearingen}, \binits{J.D.}}:
\bctitle{Interaction of vorticity with a free surface in turbulent open channel
  flow}.
In: \bbtitle{29th {{AIAA Aerospace Sciences Meeting}}},
vol. \bseriesno{1}
(\byear{1991})
\end{bchapter}
\endbibitem

%%% 26
\bibitem[\protect\citeauthoryear{Monty et~al.}{2009}]{Monty2009}
\begin{barticle}
\bauthor{\bsnm{Monty}, \binits{J.P.}},
\bauthor{\bsnm{Hutchins}, \binits{N.}},
\bauthor{\bsnm{Ng}, \binits{H.C.H.}},
\bauthor{\bsnm{Marusic}, \binits{I.}},
\bauthor{\bsnm{Chong}, \binits{M.S.}}:
\batitle{A comparison of turbulent pipe, channel and boundary layer flows}.
\bjtitle{Journal of Fluid Mechanics}
\bvolume{632},
\bfpage{431}--\blpage{442}
(\byear{2009})
\doiurl{10.1017/S0022112009007423}
\end{barticle}
\endbibitem

%%% 27
\bibitem[\protect\citeauthoryear{Moser et~al.}{1999}]{Moser1999}
\begin{barticle}
\bauthor{\bsnm{Moser}, \binits{R.D.}},
\bauthor{\bsnm{Kim}, \binits{J.}},
\bauthor{\bsnm{Mansour}, \binits{N.N.}}:
\batitle{Direct numerical simulation of turbulent channel flow up to
  {{Re}}{\textsubscript{{$\tau$}}}=590}.
\bjtitle{Physics of Fluids}
\bvolume{11}(\bissue{4}),
\bfpage{943}--\blpage{945}
(\byear{1999})
\doiurl{10.1063/1.869966}
\end{barticle}
\endbibitem

%%% 28
\bibitem[\protect\citeauthoryear{Monty et~al.}{2007}]{Monty2007}
\begin{barticle}
\bauthor{\bsnm{Monty}, \binits{J.P.}},
\bauthor{\bsnm{Stewart}, \binits{J.A.}},
\bauthor{\bsnm{Williams}, \binits{R.C.}},
\bauthor{\bsnm{Chong}, \binits{M.S.}}:
\batitle{Large-scale features in turbulent pipe and channel flows}.
\bjtitle{Journal of Fluid Mechanics}
\bvolume{589},
\bfpage{147}--\blpage{156}
(\byear{2007})
\end{barticle}
\endbibitem

%%% 29
\bibitem[\protect\citeauthoryear{Nagaosa}{1999}]{Nagaosa1999}
\begin{barticle}
\bauthor{\bsnm{Nagaosa}, \binits{R.}}:
\batitle{Direct numerical simulation of vortex structures and turbulent scalar
  transfer across a free surface in a fully developed turbulence}.
\bjtitle{Physics of Fluids}
\bvolume{11}(\bissue{6}),
\bfpage{1581}--\blpage{1595}
(\byear{1999})
\doiurl{10.1063/1.870020}
\end{barticle}
\endbibitem

%%% 30
\bibitem[\protect\citeauthoryear{Nagaosa and Handler}{2003}]{Nagaosa2003}
\begin{barticle}
\bauthor{\bsnm{Nagaosa}, \binits{R.}},
\bauthor{\bsnm{Handler}, \binits{R.A.}}:
\batitle{Statistical analysis of coherent vortices near a free surface in a
  fully developed turbulence}.
\bjtitle{Physics of Fluids}
\bvolume{15}(\bissue{2}),
\bfpage{375}--\blpage{394}
(\byear{2003})
\doiurl{10.1063/1.1533071}
\end{barticle}
\endbibitem

%%% 31
\bibitem[\protect\citeauthoryear{Nagaosa and Handler}{2012}]{Nagaosa2012}
\begin{barticle}
\bauthor{\bsnm{Nagaosa}, \binits{R.}},
\bauthor{\bsnm{Handler}, \binits{R.A.}}:
\batitle{Characteristic time scales for predicting the scalar flux at a free
  surface in turbulent open-channel flows}.
\bjtitle{AIChE Journal}
\bvolume{58}(\bissue{12}),
\bfpage{3867}--\blpage{3877}
(\byear{2012})
\doiurl{10.1002/aic.13773}
\end{barticle}
\endbibitem

%%% 32
\bibitem[\protect\citeauthoryear{Nagaosa and Saito}{1997}]{Nagaosa1997}
\begin{barticle}
\bauthor{\bsnm{Nagaosa}, \binits{R.}},
\bauthor{\bsnm{Saito}, \binits{T.}}:
\batitle{Turbulence structure and scalar transfer in stratified free-surface
  flows}.
\bjtitle{AIChE Journal}
\bvolume{43}(\bissue{10}),
\bfpage{2393}--\blpage{2404}
(\byear{1997})
\doiurl{10.1002/aic.690431003}
\end{barticle}
\endbibitem

%%% 33
\bibitem[\protect\citeauthoryear{Oberlack et~al.}{2022}]{Oberlack2022}
\begin{barticle}
\bauthor{\bsnm{Oberlack}, \binits{M.}},
\bauthor{\bsnm{Hoyas}, \binits{S.}},
\bauthor{\bsnm{Kraheberger}, \binits{S.V.}},
\bauthor{\bsnm{{Alc{\'a}ntara-{\'A}vila}}, \binits{F.}},
\bauthor{\bsnm{Laux}, \binits{J.}}:
\batitle{Turbulence {{Statistics}} of {{Arbitrary Moments}} of {{Wall-Bounded
  Shear Flows}}: {{A Symmetry Approach}}}.
\bjtitle{Physical Review Letters}
\bvolume{128}(\bissue{2}),
\bfpage{024502}
(\byear{2022})
\doiurl{10.1103/PhysRevLett.128.024502}
\end{barticle}
\endbibitem

%%% 34
\bibitem[\protect\citeauthoryear{Panton}{2009}]{Panton2009}
\begin{barticle}
\bauthor{\bsnm{Panton}, \binits{R.L.}}:
\batitle{Scaling and correlation of vorticity fluctuations in turbulent
  channels}.
\bjtitle{Physics of Fluids}
\bvolume{21}(\bissue{11}),
\bfpage{115104}
(\byear{2009})
\doiurl{10.1063/1.3249753}
\end{barticle}
\endbibitem

%%% 35
\bibitem[\protect\citeauthoryear{Pan and Banerjee}{1995}]{Pan1995}
\begin{barticle}
\bauthor{\bsnm{Pan}, \binits{Y.}},
\bauthor{\bsnm{Banerjee}, \binits{S.}}:
\batitle{A numerical study of free-surface turbulence in channel flow}.
\bjtitle{Physics of Fluids}
\bvolume{7}(\bissue{7}),
\bfpage{1649}--\blpage{1664}
(\byear{1995})
\doiurl{10.1063/1.868483}
\end{barticle}
\endbibitem

%%% 36
\bibitem[\protect\citeauthoryear{Perry et~al.}{1986}]{Perry1986}
\begin{barticle}
\bauthor{\bsnm{Perry}, \binits{a.E.}},
\bauthor{\bsnm{Henbest}, \binits{S.}},
\bauthor{\bsnm{Chong}, \binits{M.S.}}:
\batitle{A theoretical and experimental study of wall turbulence}.
\bjtitle{Journal of Fluid Mechanics}
\bvolume{165},
\bfpage{163}--\blpage{199}
(\byear{1986})
\doiurl{10.1017/S002211208600304X}
\end{barticle}
\endbibitem

%%% 37
\bibitem[\protect\citeauthoryear{Pinelli et~al.}{2022}]{Pinelli2022}
\begin{barticle}
\bauthor{\bsnm{Pinelli}, \binits{M.}},
\bauthor{\bsnm{Herlina}, \binits{H.}},
\bauthor{\bsnm{Wissink}, \binits{J.G.}},
\bauthor{\bsnm{Uhlmann}, \binits{M.}}:
\batitle{Direct numerical simulation of turbulent mass transfer at the surface
  of an open channel flow}.
\bjtitle{Journal of Fluid Mechanics}
\bvolume{933},
\bfpage{49}
(\byear{2022})
\doiurl{10.1017/jfm.2021.1080}
\end{barticle}
\endbibitem

%%% 38
\bibitem[\protect\citeauthoryear{Pirozzoli}{2023}]{Pirozzoli2023}
\begin{barticle}
\bauthor{\bsnm{Pirozzoli}, \binits{S.}}:
\batitle{Searching for the log law in open channel flow}.
\bjtitle{Journal of Fluid Mechanics}
\bvolume{971},
\bfpage{15}
(\byear{2023})
\doiurl{10.1017/jfm.2023.616}
\end{barticle}
\endbibitem

%%% 39
\bibitem[\protect\citeauthoryear{Swean et~al.}{1991}]{Swean1991}
\begin{bchapter}
\bauthor{\bsnm{Swean}, \binits{T.F.}},
\bauthor{\bsnm{Leighton}, \binits{R.I.}},
\bauthor{\bsnm{Handler}, \binits{R.A.}}:
\bctitle{Turbulence {{Modeling Near}} the {{Free Surface}} in an {{Open Channel
  Flow}}}.
In: \bbtitle{29th {{AIAA Aerospace Sciences Meeting}}}
(\byear{1991})
\end{bchapter}
\endbibitem

%%% 40
\bibitem[\protect\citeauthoryear{Scherer et~al.}{2022}]{scherer:21a}
\begin{barticle}
\bauthor{\bsnm{Scherer}, \binits{M.}},
\bauthor{\bsnm{Uhlmann}, \binits{M.}},
\bauthor{\bsnm{Kidanemariam}, \binits{A.G.}},
\bauthor{\bsnm{Krayer}, \binits{M.}}:
\batitle{On the role of turbulent streaks in generating sediment ridges}.
\bjtitle{J. Fluid Mech.}
\bvolume{930},
\bfpage{11}
(\byear{2022})
\doiurl{10.1017/jfm.2021.891}
{\href{https://arxiv.org/abs/2110.00815}{{2110.00815}}}
\end{barticle}
\endbibitem

%%% 41
\bibitem[\protect\citeauthoryear{Yao et~al.}{2022}]{Yao2022}
\begin{barticle}
\bauthor{\bsnm{Yao}, \binits{J.}},
\bauthor{\bsnm{Chen}, \binits{X.}},
\bauthor{\bsnm{Hussain}, \binits{F.}}:
\batitle{Direct numerical simulation of turbulent open channel flows at
  moderately high {{Reynolds}} numbers}.
\bjtitle{Journal of Fluid Mechanics}
\bvolume{953},
\bfpage{19}
(\byear{2022})
\doiurl{10.1017/jfm.2022.942}
\end{barticle}
\endbibitem

%%% 42
\bibitem[\protect\citeauthoryear{Zhong et~al.}{2016}]{Zhong2016}
\begin{barticle}
\bauthor{\bsnm{Zhong}, \binits{Q.}},
\bauthor{\bsnm{Chen}, \binits{Q.}},
\bauthor{\bsnm{Wang}, \binits{H.}},
\bauthor{\bsnm{Li}, \binits{D.}},
\bauthor{\bsnm{Wang}, \binits{X.}}:
\batitle{Statistical analysis of turbulent super-streamwise vortices based on
  observations of streaky structures near the free surface in the smooth open
  channel flow}.
\bjtitle{Water Resources Research}
\bvolume{52}(\bissue{5}),
\bfpage{3563}--\blpage{3578}
(\byear{2016})
\doiurl{10.1002/2015WR017728}
\end{barticle}
\endbibitem

\end{thebibliography}
% ============================================================================ %
%
%
% ============================================================================ %
\end{document}